\documentclass[12pt]{article}

\usepackage[T1]{fontenc} 
\usepackage[utf8]{inputenc}

\usepackage{amssymb,amsmath,amscd}
\usepackage{graphicx}
\usepackage[english]{babel}
\usepackage{multirow}
\usepackage{float}
\usepackage{titlesec}
\usepackage[bulletsep]{collref}
\usepackage{fullpage,epsfig,graphics,amsbsy,amssymb,cancel,slashed,mathrsfs}
\usepackage{psfrag,hyperref}
\usepackage{graphicx,color}
\usepackage{wrapfig}
\usepackage{subfigure}
\usepackage{setspace}
\usepackage{lipsum}
\usepackage{overpic}

\graphicspath{{./figures/}}

\hypersetup{
	colorlinks=true,
	linkcolor=dark-blue,
	citecolor=dark-red,
	urlcolor=dark-green,
	linktoc=page
}

\usepackage[hmarginratio=1:1,top=32mm,columnsep=20pt]{geometry} 
\usepackage{fancyhdr} 
\numberwithin{equation}{section}

\titleformat{\section}[block]{\large\bfseries\centering}{\thesection}{1em}{} 
\titleformat{\subsection}[block]{\bfseries}{\thesubsection}{1em}{} 

\usepackage{abstract}

\definecolor{dark-gray}{gray}{0.20}
\definecolor{gray}{gray}{0.30}
\definecolor{light-gray}{gray}{0.80}
\definecolor{dark-red}{rgb}{0.7,0,0}
\definecolor{dark-green}{rgb}{0.1,0.4,0}
\definecolor{dark-blue}{rgb}{0.3,0.3,0.7}
\definecolor{light-blue}{rgb}{0.8,0.8,1}
\definecolor{cardinal}{rgb}{0.6,0,0}
\definecolor{darkgreen}{rgb}{0,0.5,0}
\definecolor{golden}{rgb}{0.92, 0.7, 0}
\definecolor{midnight}{rgb}{0, 0, 0.5}
\definecolor{darkblue}{rgb}{0.2, 0, 0.8}
\definecolor{forestgreen}{rgb}{0.13, 0.55, 0.13}

\def\L{{\cal L}}

\def\R{{\cal R}}

\newcommand{\SL}{\textsl}
\def\Tr{{\rm Tr}\,}

\newcommand{\vol}{\mathrm{vol}}
\newcommand{\dd}{\mathrm{d}}
\newcommand{\e}{\mathrm{e}}
\newcommand{\f}[2]{\frac{#1}{#2}}
\newcommand{\tf}[2]{\tfrac{#1}{#2}}

\newcommand{\be}{\begin{equation}}
\newcommand{\ee}{\end{equation}}
\newcommand{\bea}{\begin{eqnarray}}
\newcommand{\eea}{\end{eqnarray}}  
\newcommand{\nn}{\nonumber}

\newcommand{\NN}{\mathcal{N}}

\newcommand{\sfrac}[2]{\mbox{$\frac{#1}{#2}$}}
\newcommand{\ms}{\!-\!}

\newcommand{\pintd}[2]{\int_{#1}^{#2}\!\!\!\!\!\!\!\!\!-\,\,\,}

\newcommand{\sign}{{\rm sign}}

\newcommand{\csch}{{\rm csch}}

\def\rmi{{\rm i}}

\def\lqft{\lambda_{\text{QFT}}}

\def\al{\alpha}

\def\eps{\epsilon}
\def\s{\sigma}

\def\SL{{\rm SL}}

\def\SO{{\rm SO}}
\def\U{{\rm U}}
\def\SU{{\rm SU}}

\title{\vspace{-30mm}\begin{flushright} \small
UUITP-32/19
\end{flushright}
\vspace{10mm}\fontsize{20pt}{24pt}\selectfont\textbf{Supersymmetric Yang-Mills, Spherical Branes, and Precision Holography}\vspace{3mm} }

\author{Nikolay Bobev$^a$, Pieter Bomans$^a$, Fri\dh rik Freyr Gautason$^{a,b}$,\\ Joseph A. Minahan$^c$ and Anton Nedelin$^d$\\[5mm]
	\normalsize $^a$Instituut voor Theoretische Fysica, K.U. Leuven\\
	\normalsize Celestijnenlaan 200D, BE-3001 Leuven, Belgium\\[1mm]
	\normalsize $^b$ University of Iceland, Science Institute\\
	\normalsize Dunhaga 3, 107 Reykjav\'ik, Iceland\\[1mm]
	\normalsize $^c$Department of Physics and Astronomy, Uppsala University\\
	\normalsize Box 516, SE-75120 Uppsala, Sweden\\[1mm]
	\normalsize $^d$Deparment of Physics, Technion\\
	\normalsize 32000 Haifa, Israel\\[3mm]
	\texttt{\small\href{mailto:nikolay.bobev@kuleuven.be}{\{nikolay.bobev}, \href{mailto:ffg@kuleuven.be}{ffg}, \href{mailto:pieter.bomans@kuleuven.be}{pieter.bomans\}@kuleuven.be}}\\
	\texttt{\small\href{mailto:joseph.minahan@physics.uu.se}{\{joseph.minahan},\href{mailto:anton.nedelin@physics.uu.se}{anton.nedelin\}@physics.uu.se}}\\
}

\date{}

\begin{document}
\maketitle

\begin{abstract}
\noindent Using supersymmetric localization we compute the free energy and BPS Wilson loop vacuum expectation values for planar maximally supersymmetric Yang-Mills theory on $S^d$ in the strong coupling limit for $2\leq d<6$. The same calculation can also be performed in supergravity using the recently found spherical brane solutions. We find excellent agreement between the two sets of results. This constitutes a non-trivial precision test of holography in a non-conformal setting.  The free energy of maximal SYM on $S^6$ diverges in the strong coupling limit which might signify the onset of little string theory. We show how this divergence can be regularized both in QFT and in supergravity.  We also consider $d=7$ with a small negative 't Hooft coupling and show that the free energy and Wilson loop vacuum expectation value agree with the results from supergravity after addressing some subtleties. 
\end{abstract}

\thispagestyle{empty}
\newpage

\setcounter{tocdepth}{2}
\pagenumbering{arabic}
\tableofcontents

\newpage

\section{Introduction}
\label{sec:intro}

Supersymmetric localization is a powerful tool to study the dynamics of strongly coupled supersymmetric QFTs which has been efficiently exploited in a variety of examples \cite{Pestun:2016zxk}. A particularly interesting application of this technique is the study of the correspondence between gauge theories and their gravity duals. In many situations the calculation of supersymmetric observables in the field theory reduces to an evaluation of a matrix integral which can then be studied in the planar limit with saddle point techniques. In the cases when the supersymmetric theory has a known gravitational dual this provides a fruitful avenue to quantitatively test the details of the AdS/CFT correspondence.

It is natural to consider questions on the interface of holography and supersymmetric localization for conformal theories with maximal supersymmetry, like four-dimensional $\mathcal{N}=4$ SYM and the three-dimensional ABJM theory, on the round sphere. Indeed this was pursued extensively and many important developments are summarized in \cite{Pestun:2016zxk}. These two examples also offer the possibility to break conformal invariance and part of the supersymmetry while still maintaining calculational control both in the field theory \cite{Russo:2012ay,Buchel:2013id,Russo:2013qaa,Russo:2013kea} and the supergravity side \cite{Freedman:2013ryh,Bobev:2013cja,Bobev:2016nua,Bobev:2018hbq,Bobev:2018wbt,Kim:2019rwd,Kim:2019ewv,Kim:2019feb}. This collection of results provides a non-trivial precision test of holography away from the conformal limit. Our goal in this paper is to extend this success to other non-conformal theories with maximal supersymmetry arising from string theory.

The theories we consider are maximally supersymmetric gauge theories on the round sphere, $S^d$. In dimension $2\leq d\leq7$ these theories are not conformal for $d\neq4$ and admit a Lagrangian which preserves 16 supercharges \cite{Blau:2000xg,Minahan:2015jta}. Supersymmetric localization reduces the path integral of the theory to an ordinary matrix integral. Despite this drastic simplification the explicit evaluation of this integral is still non-trivial due to the presence of non-perturbative effects like instantons. When the rank of the gauge group is large it is believed that these non-perturbative effects are suppressed and the matrix integral becomes more tractable. As we discuss in detail below, for all values of $d$ it is possible to compute the free energy and the vacuum expectation values (VEV) of a supersymmetric Wilson loop using this matrix model.\footnote{See \cite{Giombi:2014xxa} for calculations of the free energy on $S^d$ of QFTs without gauge fields.} A further simplification occurs in the limit where the dimensionless 't Hooft coupling, defined as
\be\label{tHooft}
\lambda\equiv \R^{d-4}g_{\text{YM}}^2N
\ee
 where $\R$ is the radius of $S^d$, is large.  In this case the results can be written in analytic form and can be formally analytically continued even to non-integer values of $d$.

The gravity dual of these maximally supersymmetric Yang-Mills theories (MSYM) on flat space is given by the near horizon geometry of the D$p$-brane solutions in supergravity with $d=p+1$ \cite{Itzhaki:1998dd}. To study the MSYM theories on $S^d$ one needs a generalization of these solutions to D$p$-branes with spherical worldvolume. Indeed, such spherical brane solutions exist and were constructed explicitly in \cite{Bobev:2018ugk}.\footnote{See also \cite{Triendl:2015lka,Minasian:2017rgh,Minasian:2017pez,Maxfield:2015evr} for other constructions of supersymmetric solutions sourced by curved Euclidean branes.} Equipped with these supergravity backgrounds we can apply the tools of holography and compute the free energy and Wilson loop VEV at large $\lambda$. The holographic free energy is calculated by evaluating the on-shell action of the supergravity solution while the Wilson loop VEV is computed by first finding an appropriately embedded probe string and then computing the Nambu-Goto action on-shell. Both of these calculations can be performed explicitly and the results are in agreement with the ones obtained by supersymmetric localization.

We encounter several subtleties in our calculations. In the supersymmetric localization analysis the large $N$ limit of the matrix model admits a simple saddle point evaluation only for $3<d<6$. For values of $d$ outside of this range we have to perform a careful analytic continuation.  For the case of $d=3$, one would naively expect that there would be no dependence on $\lambda$ since the Yang-Mills action is $Q$-exact in three dimensions.  However, the contribution from the localization determinant  diverges for $d=3$ with $\NN=8$ supersymmetry, offsetting the $Q$-exactness of the action.  By setting $d= 3+\eps$ and sending $\eps$ to zero we find that the free-energy is indeed independent of $\lambda$, but the  Wilson loop VEV depends nontrivially on $\lambda$.  We then show that these results can be reproduced in supergravity at large $\lambda$, including nontrivial pre-factors.  While the strong coupling results can be obtained by analytically continuing the results found for $3<d<6$, we can actually do more and find the  Wilson loop VEV in terms of a simple function of $\lambda$ which is valid for all values of the coupling.

For $d=2$ another subtlety arises.   The standard extension of a ${\cal N}=(2,2)$ vector multiplet on the sphere is $Q$-exact \cite{Benini:2012ui,Doroud:2012xw}, but this action cannot be extended to 16 supersymmetries by adding extra fields.  However, there is another action that preserves supersymmetry that can be extended and is {\it not} $Q$-exact.  This then leads to nontrivial dependence on $\lambda$.   Again we can analytically continue our results down to $d=2$ to find the free energy and the Wilson loop VEV.   We show that  supergravity reproduces the Wilson loop VEV and, with an appropriate counterterm, can also reproduce the free energy,

At $d=5$ we reproduce previous results from the literature for the free energy and Wilson loop \cite{Kim:2012ava,Kallen:2012zn,Minahan:2013jwa}.  In this case there is a well-known mismatch between the free energy coming from localization and that coming from the on-shell action of the M theory dual of the six-dimensional  $(2,0)$ theory with one direction compactified on a circle.  In this paper we consider the IIA supergravity dual directly and show that one can add  counterterms which is allowed because of the partial breaking of the $R$-symmetry and which can cancel the mismatch.  This is reminiscent of the difficulties encountered in \cite{Genolini:2016sxe,Genolini:2016ecx} in the context of holographic renormalization for AdS$_{5}$ with an $S^3\times S^1$ boundary which ultimately lead to the introduction of non-covariant counterterms. 

The cases $d=6$ and $d=7$ are particularly subtle due to the appearance of divergences in the matrix model. For $d=6$ the divergence appears to be severe and perhaps signals the onset of the $(1,1)$ little string theory which is the UV completion of maximal SYM in six dimensions. Nevertheless, we find a regularization procedure of the matrix model which leads to finite results for both the free energy and the Wilson loop VEV. 

For $d=7$ we again observe a divergence in the matrix model which can be handled using a more standard UV regularization. At weak 't Hooft coupling the matrix model is similar to the lower dimensional cases.  As we increase the regularized $\lambda$, or equivalently decrease $\lambda^{-1}$, one finds that we can smoothly continue $\lambda^{-1}$ through zero and take it to large negative values.   It is in this regime with small negative 't Hooft coupling that we can compare to supergravity, where we find a match for both the free energy and the Wilson loop VEV.   This fits nicely with an observation made by Peet and Polchinski  \cite{Peet:1998wn} who speculated that there were two weakly coupled theories in seven dimensions, the usual weakly coupled supersymmetric gauge theory and some other weakly coupled theory that is described by supergravity.  Here we see that the supergravity dual is still a gauge theory, but with a flipped sign for the coupling.  Furthermore, since the coupling is weak, albeit negative, the saddle point is sharply peaked, even for finite $N$.  This parallels the observation in \cite{Itzhaki:1998dd} that the supergravity description can be trusted even for small $N$.

The analysis on the gravity side for all $d\neq4$  goes beyond the realm of the usual holographic dictionary. The spherical brane solutions for $d\neq4$ are not asymptotically locally AdS and therefore there is no generally established holographic renormalization procedure. Despite this obstacle we are able to adapt the results in \cite{Kanitscheider:2008kd,Kanitscheider:2009as} to our setting and construct appropriate counterterms in supergravity which lead to a finite on-shell action for the spherical brane backgrounds and the probe strings. The approach of \cite{Kanitscheider:2008kd,Kanitscheider:2009as} is however not applicable for $d=6$ due to the linear dilaton characteristic of the little string theory. Inspired by the regularization procedure in the matrix model analysis and the results in \cite{Cotrone:2007qa,Marolf:2007ys,Aharony:2019zsx} we are able to propose a way to cancel the divergences appearing in the spherical D5-brane solution and obtain an agreement with the results from supersymmetric localization.

In the next section we summarize the maximally supersymmetric Yang-Mills theory on $S^d$ and show how to compute its free energy and the VEV of a BPS Wilson loop using supersymmetric localization. In Section~\ref{sec:sugra} we summarize the spherical brane solutions and the holographic renormalization procedure we use. Section~\ref{sec:FandWLcases} is devoted to a cases by cases analyses of the QFT and supergravity evaluation of the free energy and the Wilson loop VEV for $2\leq d\leq7$. We conclude in Section~\ref{sec:discussion} with a short discussion. In the appendices we summarize and further explain many technical results used throughout the paper.

\section{Field theory and supersymmetric localization}
\label{sec:fieldtheory}

The $d$-dimensional maximally supersymmetric Yang-Mills theory (MSYM) can be put on the round sphere $S^d$ while preserving all 16 supercharges. If $d\neq 4$ then MSYM  is not superconformal and the fact that one can place the theory on a sphere and still preserve supersymmetry is non-trivial and can be done only for $d\leq7$, see \cite{Blau:2000xg} and
\cite{Minahan:2015jta}. The curvature of the sphere induces new couplings in the MSYM action which break the $\SO(1,9-d)$ R-symmetry of the theory in flat Euclidean space to $\SU(1,1)\times \SO(7-d)$. One advantage of placing  MSYM  on a sphere is that one can employ the powerful techniques of supersymmetric localization to calculate  certain physical observables exactly,  see \cite{Pestun:2016zxk} for a review.  This was pursued in \cite{Minahan:2015jta,Minahan:2015any,Gorantis:2017vzz} and we summarize and extend these results below.

\subsection{Localization for MSYM on $S^d$}
\label{section:localisation}

Our starting point is the MSYM Lagrangian on $S^d$ with radius $\R$, which is given by\footnote{Here we are replacing the Yang-Mills coupling $g_{\rm  YM}^2$ in  \cite{Minahan:2015any} by $2g_{\rm YM}^2$  to match the conventions used in supergravity.} \cite{Blau:2000xg,Minahan:2015jta}
\be\label{Lss}
\begin{split}
\L&=-\frac{1}{2g_{\rm  YM}^2}\Tr\Bigg(\frac12F_{MN}F^{MN}-\Psi\slashed{D}\Psi+\frac{(d-4)}{2\R}\Psi\Lambda\Psi+\frac{2(d-3)}{\R^2} \phi^A\phi_A\\
&\hspace{3cm}+\frac{(d-2)}{\R^2}\phi^i\phi_i +\frac{2\,\rmi}{3\R}(d-4)[\phi^A,\phi^B]\phi^C\varepsilon_{ABC}-K_mK^m\Bigg)\,.
\end{split}
\ee
The indices $M,N=0,\dots 9$  arise from dimensional reduction of ten-dimensional super Yang-Mills.  In the reduction the ten-dimensional gauge field divides into a $d$-dimensional gauge field and $10-d$ scalar fields. Accordingly, the $M,N$ indices are broken up into the coordinate indices on $S^d$, $\mu,\nu=1,\dots d$, and scalar indices $I,J=0,d+1,\dots 9$.  The scalar indices  themselves split further into indices $A,B=0,8,9$ and $i,j=d+1,\dots 7$. The field-strengths with components along the scalar dimensions are $F_{\mu I}=D_\mu \phi_I$ and $F_{IJ}=-\rmi[\phi_I,\phi_J]$.  The scalar field $\phi_0$ originates from the  time-like component  of the ten-dimensional gauge field, and so  has a wrong-sign kinetic term. The $\Psi$ are 16 component real  chiral spinors satisfying $\Gamma^{11}\Psi=\Psi$ and we have defined $\Lambda=\Gamma^{089}$.    There are also 7 auxiliary fields $K_m$ which allow for an off-shell formulation of supersymmetry. 

The terms in the action proportional to $\mathcal{R}^{-1}$ and $\mathcal{R}^{-2}$ break the $R$-symmetry from $\SO(1,9-d)$ to $\SU(1,1)\times \SO(7-d)$, except for $d=4$ and $d=7$.  Note that the Lagrangian $\L$ is obtained as a deformation of the dimensional reduction of the ten-dimensional SYM Lagrangian in Lorentzian signature and we have not Wick rotated the ten-dimensional time coordinate. 

The Lagrangian in \eqref{Lss} is invariant under the off-shell supersymmetry transformations
\bea\label{susyos}
\delta_\eps A_M&=&\eps\,\Gamma_M\Psi\,,\nn\\
\delta_\eps \Psi&=&\frac12 \Gamma^{MN}F_{MN}\eps + \frac{2(d-3)}{d}\Gamma^{\mu A}\phi_A\nabla_\mu\,\eps  +\frac{2}{d}\Gamma^{\mu i}\phi_i\nabla_\mu\,\eps + K^m\nu_m\, ,\nn\\
\delta_\eps K^m&=&-\nu^m\slashed{D}\Psi+\frac{(d-4)}{2\R}\nu^m\Lambda\Psi\,,
\eea
where $\eps$  is a bosonic 16 component real chiral spinor that satisfies the conformal Killing spinor equation  
\be\label{cks}
\nabla_\mu\eps=\frac{1}{2\R}\Gamma_\mu\Lambda\eps\,.
\ee
The $\nu^m$ are seven commuting spinors that satisfy $\nu^m\Gamma^M\eps=0$, $\nu^m\Gamma^M\nu^n=\delta^{nm}\eps\Gamma^M\eps$ \cite{Pestun:2007rz,Minahan:2015jta}.   

The theory with Lagrangian \eqref{Lss}  can be localized using a particular supercharge \cite{Pestun:2007rz,Minahan:2015jta}.  Given any $\eps$ satisfying \eqref{cks} we can define a vector field $v^M\equiv\eps\Gamma^M\eps$ that automatically satisfies $v_Mv^M=0$.
We then choose $\eps$ so that $v^0=1$, $v^{8,9}=v^i=0$, and along one particular equator of the sphere $v_\mu v^\mu=1$.  We will later take the large $N$ limit where it is assumed that instantons can be ignored \cite{Russo:2012ay,Bobev:2013cja}.  In this situation the theory localizes onto the locus where $A_\mu=0$, $\phi_I=0$ for $I\ne0$,  $\nabla_\mu\phi^0=0$, and 
$K_m=-\frac{(d-3)}{\R}\phi_0(\nu_m\Lambda\eps)$.   Wick rotating the time direction leads to the transformations $\L\to-\rmi\L$, $\phi^0\to \rmi\phi^0$, and $K^m\to \rmi K^m$.  After defining a dimensionless $N\times N$ Hermitian matrix $\s\equiv\R\phi^0$, the partition function for general $d$ reduces to \cite{Minahan:2015jta,Minahan:2015any,Gorantis:2017vzz}
\bea
Z&=&\int\limits_{\rm Cartan} [d\s]~\exp\left(-  \frac{4\pi^{\frac{d+1}{2}}\mathcal{R}^{d-4}}{g_{\rm YM}^2\Gamma\left(\frac{d-3}{2}\right)}\Tr\,\sigma^2\right) Z_{\rm 1-loop}(\s)  + \mbox{instantons}\,.\label{partfun}
\eea
 $Z_{\rm 1-loop}(\s)$ is the contribution of the Gaussian fluctuations about the localized fixed point, and when combined with the Vandermonde determinant  is given by
\be\label{16susydet}
Z_{\rm 1-loop} (\s)\prod_{\gamma>0}\langle\gamma,\sigma\rangle^2=\prod_{\gamma>0
}\prod_{n=0}^\infty\left(\frac{n^2+\langle\gamma,\sigma\rangle^2}{(n+d-3)^2+\langle\gamma,\sigma\rangle^2}\right)^{\frac{\Gamma(n+d-3)}{\Gamma(n+1)\Gamma(d-3)}}\,,
\ee
where $\gamma$ are the positive roots for the gauge group.    If $d<6$ then \eqref{16susydet} is convergent.  For $d\ge6$ it diverges and will need to be regularized.  For the rest of this section we  assume that $d<6$.  The $d=6$ and $d=7$ cases will be considered separately. Notice that in the matrix model defined by \eqref{partfun}, the integration over $\sigma$ is restricted to adjoint matrices in the Cartan of the gauge group. We can therefore fully parametrize $\sigma$ by its eigenvalues $\sigma_i$.

\begin{figure}[!btp]
	\centering
	{\includegraphics[width=.9\linewidth]{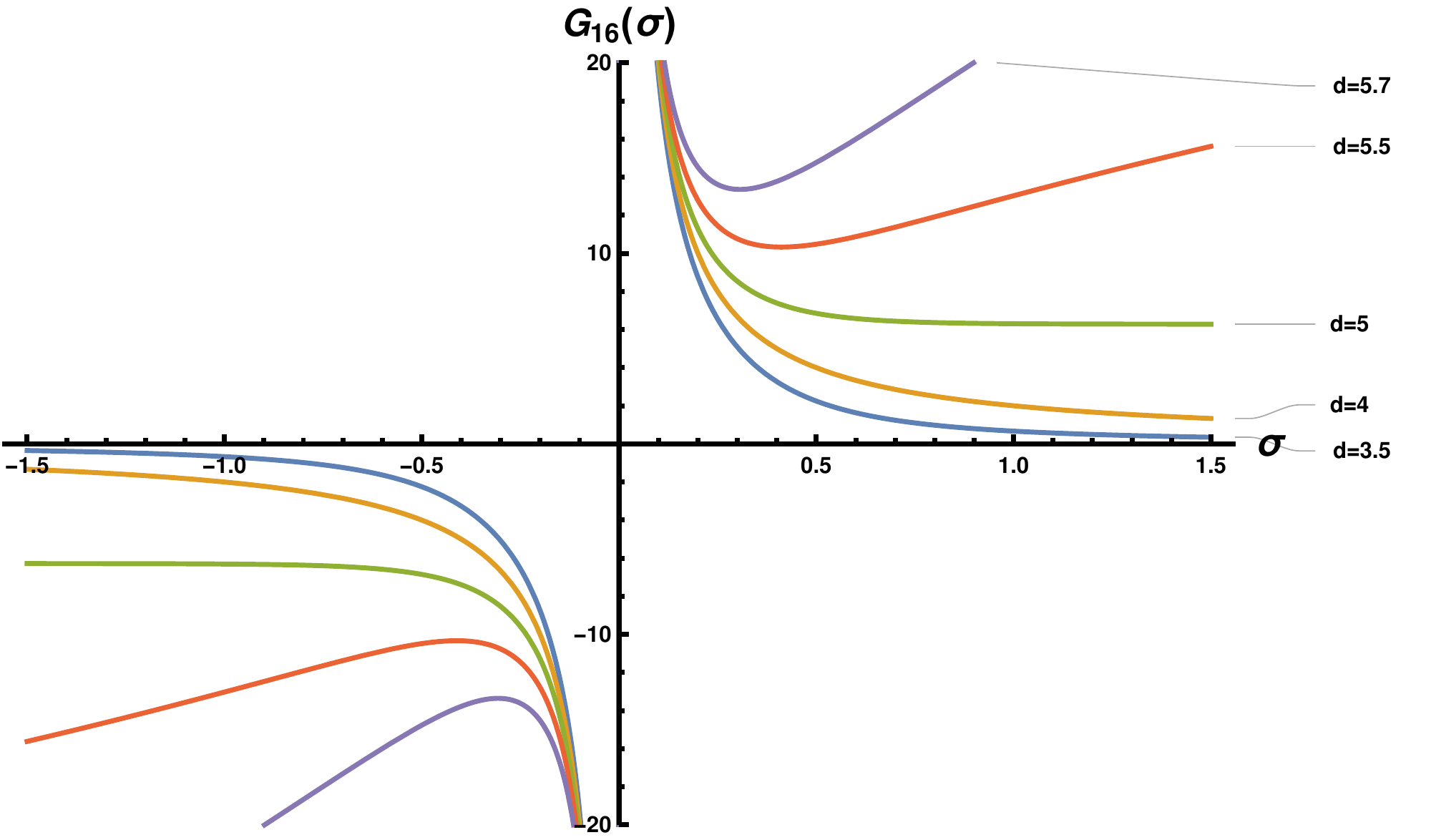}}
	\caption{The kernel $G_{16}(\s)$ for various values of $d$.  For $|\s|\ll1$ the curves approach the same weak coupling behavior.  For $|\s|>1$ they approach different strong coupling behavior. 
	}
	\label{kernel}
\end{figure}

We now take the large $N$ limit and drop the instanton contributions.   The partition function is now dominated by a saddle point   whose equations are given by  
\be\label{saddlept}
\frac{C_1 N}{\lambda}\,\sigma_i=\sum_{j\ne i} G_{16}(\s_{ij})\,,\qquad C_1\equiv\frac{8\pi^{\frac{d+1}{2}}}{\Gamma\left(\tfrac{d-3}{2}\right)}\,,
\ee
where $\lambda$ is the dimensionless 't Hooft coupling defined in \eqref{tHooft} and $\s_{ij}\equiv\s_i-\s_j$. 
The kernel $G_{16}(\s)$ is given by \cite{Minahan:2015any}
\be\label{G16}
\frac{\rmi G_{16}(\s)}{\Gamma(4-d)}=\frac{\Gamma(-\rmi\,\s)}{\Gamma(4\ms d-\rmi\,\s)}-\frac{\Gamma(\rmi\,\s)}{\Gamma(4\ms d+\rmi\,\s)}-\frac{\Gamma(d\ms3-\rmi\,\s)}{\Gamma(1-\rmi\,\s)}+\frac{\Gamma(d\ms3+\rmi\,\s)}{\Gamma(1+\rmi\,\s)}\,.
\ee
The behavior of the kernel $G_{16}(\sigma)$ is shown in Figure~\ref{kernel} for various values of $d$. Notice that in the figure we are not restricting the dimension $d$ to be an integer. Indeed the kernel $G_{16}(\sigma)$ is a meromorphic function of $d$. 

For small eigenvalue separations where $|\s_{ij}|\ll1$, the kernel has the weak coupling behavior
\be\label{weakker}
G_{16}(\s_{ij})\approx\frac{2}{\s_{ij}}\,,
\ee
which is independent of $d$. However, we are interested in strongly coupled theories where $\lambda\gg1$.  In this case the central potential for the eigenvalues is relatively weak so the repulsive force coming from the kernel pushes the eigenvalues far apart for $d<6$.  Hence, for generic $i$ and $j$ we have that $|\s_{ij}|\gg1$.  In this range  \eqref{G16} is approximately
\be\label{saddlestrong}
G_{16}(\s_{ij})\approx C_2|\s_{ij}|^{d-5}\sign(\s_{ij})\,,
\ee
where
\be\label{C2def}
C_2=2(d-3)\Gamma(5\ms d)\sin\tfrac{\pi(d-3)}{2}\,.
\ee
The saddle point equation then becomes
\be
\frac{C_1}{\lambda}N\s_i=C_2\sum_{j\ne i}|\s_i-\s_j|^{d-5}\sign(\s_i-\s_j)\,.
\label{eom:strong}
\ee
Notice that $C_2$ in \eqref{C2def}   has a pole at  $d=6$ and a double zero at $d=3$. This restricts our general analysis to the range $3<d<6$. We will return  to $d=2,3$ in Section~\ref{sec:FandWLcases}.

We next define the eigenvalue density $\rho(\s)$,
\be\label{densdelta}
\rho(\sigma)\equiv N^{-1}\sum_{i=1}^N\delta(\s-\s_i)\,.
\ee
Assuming strong coupling, the saddle point equation \eqref{eom:strong} for $3<d<6$  becomes 
\be\label{eom}
\frac{C_1}{\lambda}\,\s=C_2\pintd{-b}{b}d\s'\rho(\s')|\s-\s'|^{d-5}\sign(\s-\s')\,,
\ee
where $ b$, given below,  sets the endpoints of the eigenvalue distribution.
Taking the large $N$ limit and using the result in \eqref{intstr}, we see that \eqref{eom} is satisfied if the density has the form
\be\label{eigdens}
\rho(\s)=\frac{C_1\sin\frac{\pi(d-1)}{2}}{\pi\lambda C_2(d-5)(b^2-{\s}^2)^{(d-5)/2}} = \frac{2\pi^{\frac{d+1}{2}}}{\pi\lambda \Gamma(6-d)\Gamma(\tfrac{d-1}{2})(b^2-{\s}^2)^{(d-5)/2}}\,.
\ee
Using \eqref{rhoint}, we can properly normalize the density by setting
the eigenvalue endpoint $b$ to
\bea\label{bres}
b = (4\pi)^{\frac{d+1}{2(d-6)}}\Big( 32\lambda\Gamma\big(\tfrac{8-d}{2}\big)\Gamma\big(\tfrac{6-d}{2}\big)\Gamma\big(\tfrac{d-1}{2}\big)\Big)^{\frac{1}{6-d}}\,.
\eea

\begin{figure}[!btp]
	\centering
	\subfigure[$d=4.5$, $N=80$, $\lambda=350$]{\includegraphics[width=0.32\linewidth]{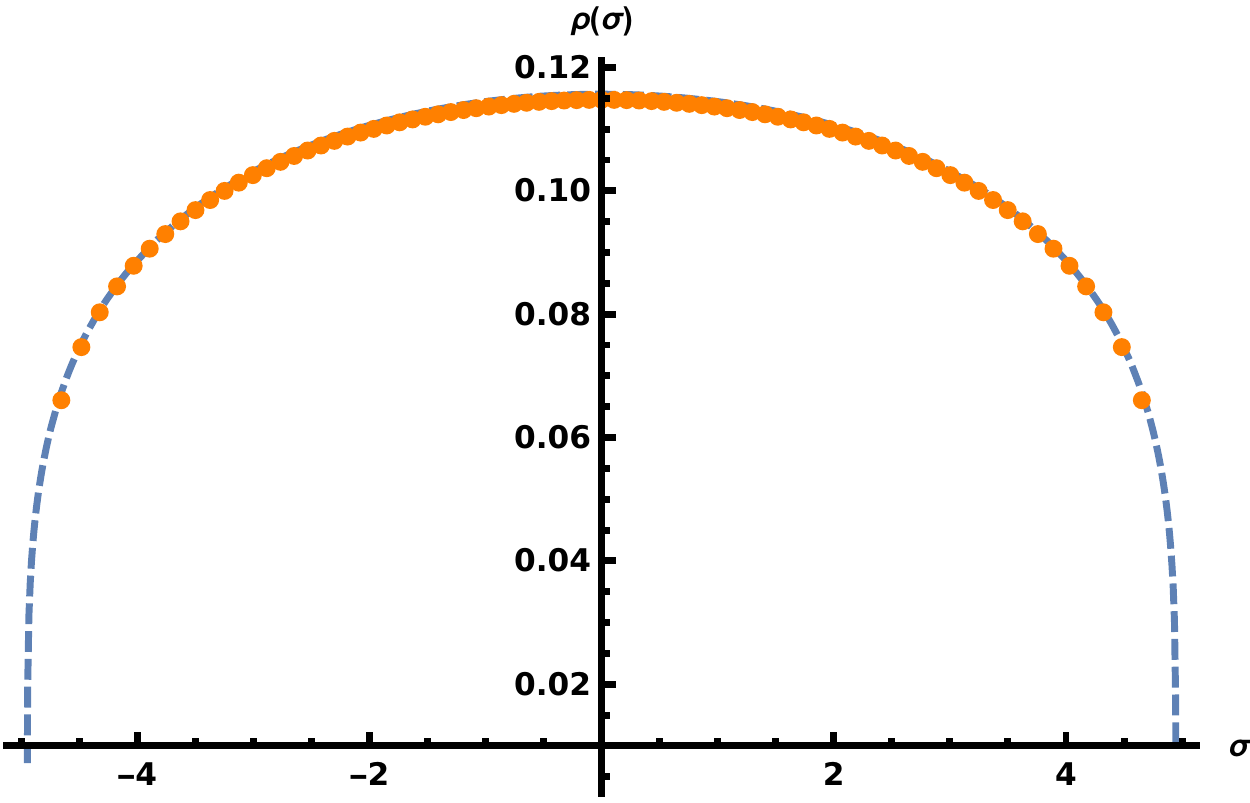}}
	\subfigure[$d=4.98$, $N=100$, $\lambda=500$]{\includegraphics[width=0.32\linewidth]{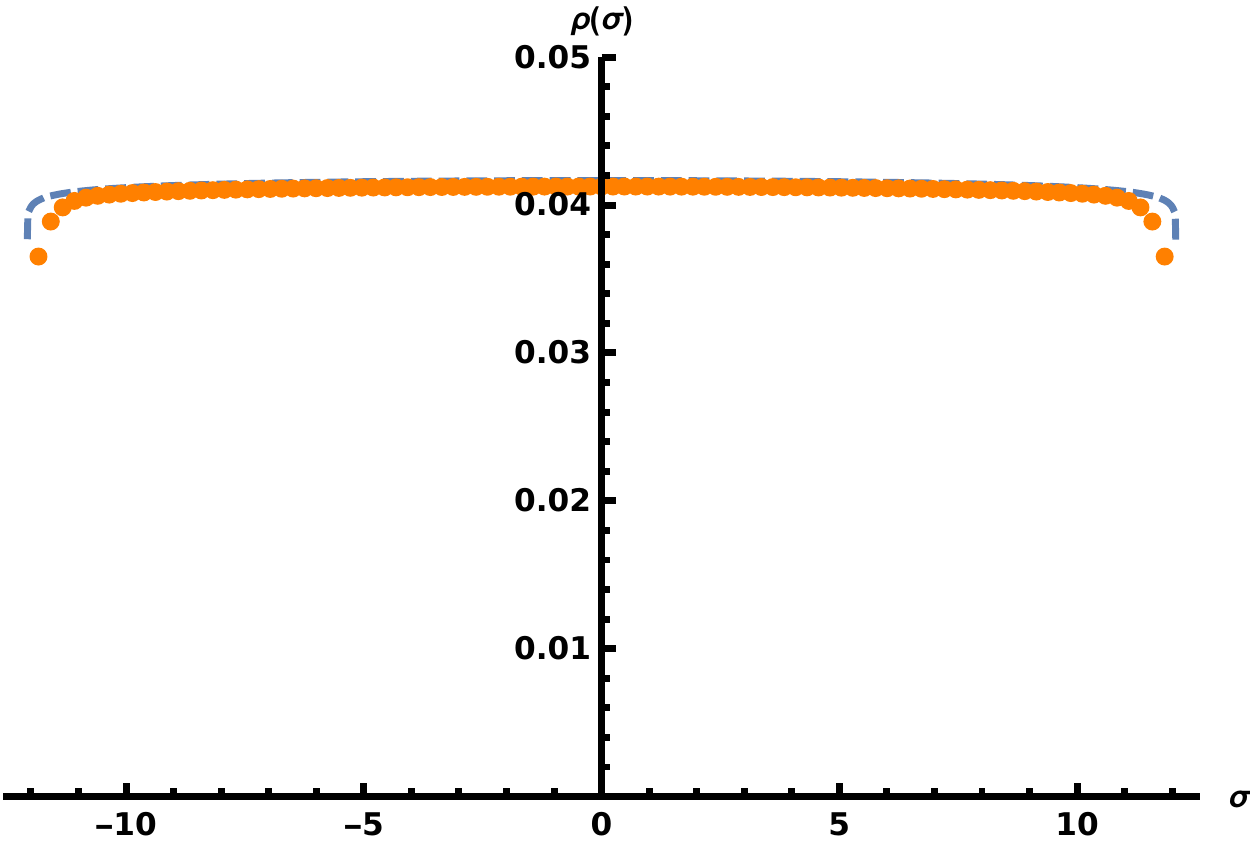}}   
	\subfigure[$d=5.5$, $N=80$, $\lambda=100$]{\includegraphics[width=0.32\linewidth]{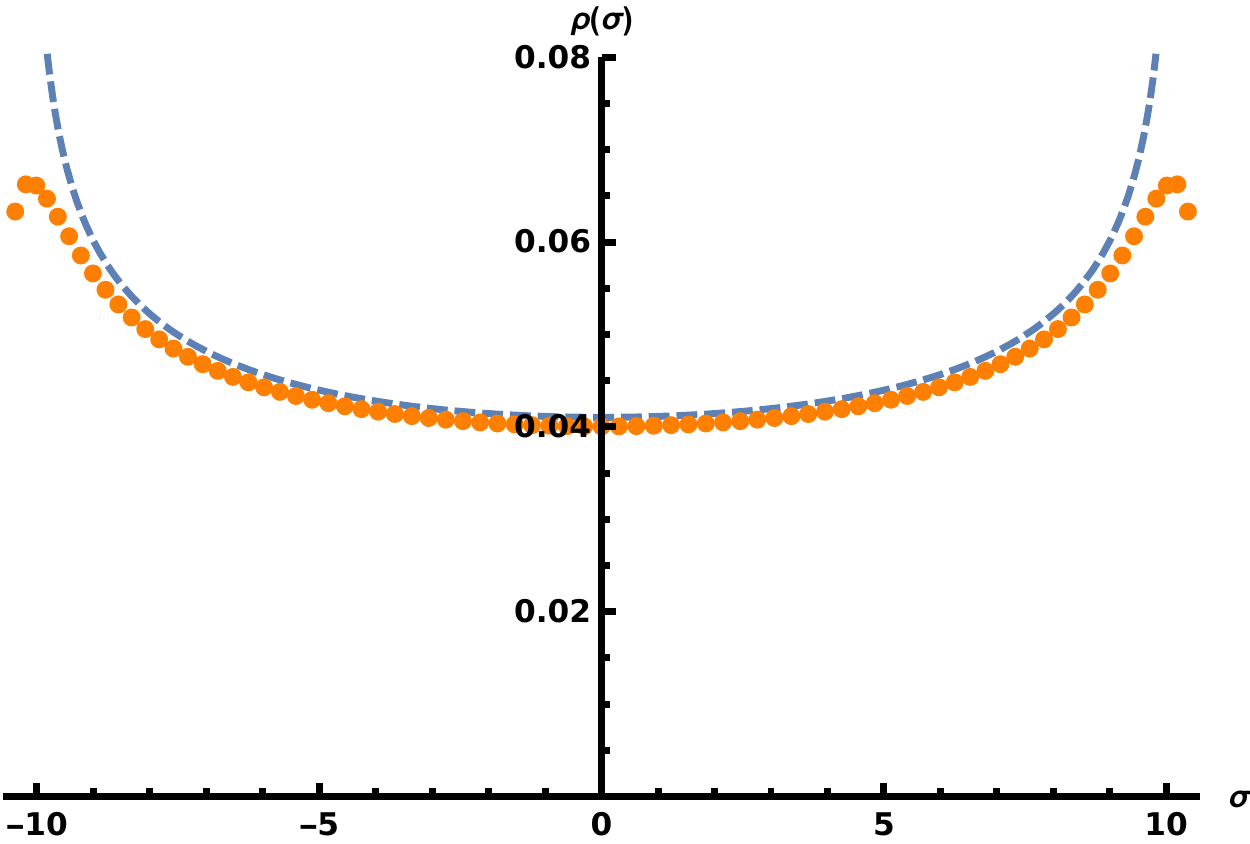}}
	\caption{The eigenvalue density obtained from the numerical 
		solutions of the full saddle point equations (\ref{saddlept})  with various choices of parameters. The dashed lines represent the eigenvalue density in (\ref{eigdens}). 
	}
	\label{pic:density}
\end{figure}

To verify  the validity of the strong coupling approximation in \eqref{saddlestrong} we can test the solutions to the saddle point equation numerically using the full function $G_{16}(\s_{ij})$ defined in \eqref{G16}. As can be seen from the graphs in Figure~\ref{pic:density}, the numerical solutions at strong coupling are in very good agreement with the eigenvalue density \eqref{eigdens} in dimensions $3<d<6$.  

\subsection{The free energy and the BPS Wilson loop VEV from localization}
\label{section:FE}

In the strong coupling regime the large $N$ limit of the free energy, $F=-\log Z$, is given by
\be\label{FEsc}
F=N^2\left(\frac{C_1}{2\lambda}\int_{-b}^b d\s\rho(\s)\s^2-\frac{C_2}{2(d-4)}\int_{-b}^b d\s\rho(\s)\int_{-b}^b d\s'\rho(\s')|\s-\s'|^{d-4}\right)\,.
\ee
Dividing through by the $N^2$ factor and performing the second integral over $\s$ by parts gives
\bea
\frac{F}{N^2}&=&\frac{C_1}{2\lambda}\int_{-b}^b d\s\rho(\s)\s^2-\frac{C_2f(b)}{d-4}\int_{-b}^b d\s'\rho(\s')|b-\s'|^{d-4}\nn\\
&&\qquad\qquad\qquad\qquad+\frac{C_2}{2}\int_{-b}^b d\s f(\s)\int_{-b}^b d\s'\rho(\s')|\s-\s'|^{d-5}\,,
\eea
where $f(\s)$ is defined in \eqref{fdef} and we used the fact that it is an odd function.  Using \eqref{eom} in the last integral and integrating by parts we find 
\be
\frac{F}{N^2}=\frac{C_1}{4\lambda}\int_{-b}^b d\s\rho(\s)\s^2+\frac{C_1}{2\lambda}f(b)b^2-\frac{C_2f(b)}{d-4}\int_{-b}^b d\s'\rho(\s')|b-\s'|^{d-4}\,.
\ee
The remaining integrals are evaluated in \eqref{I_1def} and \eqref{I_2def}.  Using these, as well as $f(b)=1/2$ and the expression for $b$ in \eqref{bres}, we can simplify the free energy to 
\bea\label{FreeEn}
\frac{F}{N^2}&=&-\frac{C_1}{2\lambda}\frac{(6-d)}{(8-d)(d-4)}\,b^2\nn\\
&=&-\frac{ 16\pi^{\frac{(d+1)(4-d)}{2(6-d)}}(6-d)}{\lambda\,\Gamma(\frac{d-3}{2})(8\ms d)(d\ms4)}
\Big( \tfrac{\lambda}{4}\Gamma\big(\tfrac{8-d}{2}\big)\Gamma\big(\tfrac{6-d}{2}\big)\Gamma\big(\tfrac{d-1}{2}\big)\Big)^{\frac{2}{6-d}}\,.
\eea
This is our final result for the free energy as a function of $d$ in the strong coupling limit.  

A $\frac12$-BPS Wilson loop $W$  wraps the equator of $S^d$ and has a VEV given by
\be
\langle W\rangle=\Big\langle\Tr\Big( {\rm\bf P}\e^{\rmi\oint dx^\mu A_\mu+\rmi\oint ds\, n_A \phi^A}\Big)\Big\rangle
\ee
where $n_A n^A=1$ and $n_A$ is fixed in its direction.  If the loop is chosen to be invariant with respect to the same supersymmetry used to localize the partition function then the Wilson loop can also be localized.  For our choice of supersymmetry this sets $n_0=1$ \cite{Pestun:2007rz,Minahan:2015jta}
and  in the large $N$ limit the Wilson loop becomes
\be\label{WL}
\langle W\rangle=\Big\langle\Tr\Big( {\rm\bf P}\e^{\rmi\oint ds\cdot \phi_0}\Big)\Big\rangle \approx \int_{-b}^bd\s\rho({\s})\e^{2\pi\s}=(\pi b)^{\frac{d-6}{2}}\Gamma\left(\sfrac{8-d}{2}\right)I_{\frac{6-d}{2}}(2\pi b)\,,
\ee
where we used the eigenvalue density in \eqref{eigdens} to evaluate the integral.  The $I_{\frac{6-d}{2}}(2\pi b)$ are  modified Bessel functions which reduce to spherical Bessel functions when $d$ is odd.

The result in \eqref{WL} is  valid for any value of $\lambda$  in $d=4$.   In section 4 we will show that this is also true for $d=3$.  For all other $d$ the result in \eqref{WL} is valid only for large $\lambda$.  
 In comparing to supergravity  we will be mainly interested in the strong coupling limit  anyway.   In this case the Wilson loop VEV is generally determined by the highest eigenvalue $b$, where we find
\be\label{WLstr}
\langle W\rangle\sim \e^{2\pi b}\,.
\ee

In the next section we discuss how one can obtain these results for the free energy and the Wilson loop VEV from supergravity.

\section{Supergravity}
\label{sec:sugra}

In this section we summarize the spherical D$p$-brane type II supergravity solutions found in \cite{Bobev:2018ugk}. These solutions are expected to provide a holographic dual to the MSYM theories on $S^d$ discussed above. Note that we use $p$ and $d=p+1$ interchangeably throughout the rest of this paper.  We then present a roadmap to computing the holographic free energy and $\tf12$-BPS Wilson loops VEV using these supergravity solutions. The explicit comparison between field theory and supergravity will be carried out in Section~\ref{sec:FandWLcases}.

\subsection{Spherical branes}
\label{subsec:sphericalbranes}

In \cite{Bobev:2018ugk} type II supergravity solutions preserving sixteen supercharges, corresponding to the backreaction of D$p$-branes with a spherical worldvolume, were constructed. These backgrounds are found by starting with $(p+2)$-dimensional maximal gauged supergravity and subsequently lifting the solutions up to type IIA/B supergravity. A short discussion of the gauged supergravity construction can be found in Appendix~\ref{app:sugraspheres}, see \cite{Bobev:2018ugk} for more details. The  type II string frame metric for these backgrounds is given by\footnote{In this paper we use  $\eta$ to denote the scalar $\lambda$ in \cite{Bobev:2018ugk}.}
\be\label{10dmetricgeneral}
\dd s_{10}^2 = \f{\e^{\eta}}{\sqrt{\mathstrut Q}}\left(\dd s_{p+2}^2 + \f{\e^{\f{2(p-3)}{6-p}\eta}}{g^2}\left(\dd \theta^2 + P\cos^2\theta~\dd \widetilde{\Omega}_2^2 + Q\sin^2\theta~\dd \Omega_{5-p}^2\right)\right)\,.
\ee
Here $g$ is the gauge coupling of the $(p+2)$-dimensional supergravity theory and can be related to the ten-dimensional string theory constants as
\begin{equation}\label{eq:defg}
(2\pi \ell_s g)^{p-7} = \frac{g_s N}{2\pi V_{6-p}}\,,
\end{equation}
where $N$ is the number of D$p$-branes, $g_s$ is the string coupling, $\ell_s$ is the string length, and $V_{n} = 2\pi^{(n+1)/2}/\Gamma(\tfrac{n+1}{2})$ is the volume of the unit radius $n$-sphere. In \eqref{10dmetricgeneral} $\dd\Omega_{5-p}^2$ is the metric on the unit radius $(5-p)$-sphere, and $\dd\widetilde{\Omega}_2^2$ is the metric on the unit radius two-dimensional de Sitter space. Together with the coordinate $\theta$ these form a squashed $(8-p)$-dimensional de Sitter space. The $(p+2)$-dimensional factor of the metric, $\dd s_{p+2}^2$, is given by
\be\label{pp2metric}
ds_{p+2}^2 = \dd r^2 + \e^{2A(r)}\dd\Omega_{p+1}^2\,,
\ee
and $\dd \Omega_{p+1}^2$ is the metric on the round $(p+1)$-sphere wrapped by the D$p$-branes. The function $A(r)$ is determined in terms of the scalars $\eta(r)$, $X(r)$, and $Y(r)$ by an algebraic equation as shown in Appendix \ref{app:sugraspheres}. The squashing functions $P$ and $Q$ are determined in terms of the gauged supergravity scalars as
\bea\label{eq:PQdef}
P &=& \left\{\begin{array}{ll} X\big(X\sin^2\theta+(X^2-Y^2)\cos^2\theta\big)^{-1} & \text{for $p<3$}\,,\\
	X\big(\cos^2\theta+X\sin^2\theta\big)^{-1} & \text{for $p>3$}\,,\end{array}\right.\\
Q &=& \left\{\begin{array}{ll} X\big(\sin^2\theta+X\cos^2\theta\big)^{-1} & \text{for $p<3$}\,,\\
	X\big(X\cos^2\theta+(X^2-Y^2)\sin^2\theta\big)^{-1} & \text{for $p>3$}\,.\end{array}\right.
\eea
The $\dd\Omega_{p+1}^2$ and $\dd\Omega_{5-p}^2$ factors in the metric realize the $\SO(p+2)\times \SO(6-p)$ spacetime and compact R-symmetries of the maximal SYM theory on the $(p+1)$-sphere. The non-compact $\SU(1,1)$ factor of the R-symmetry group on the other hand is realized as the isometry group of the two-dimensional de Sitter space with metric
\be
\dd\widetilde{\Omega}_2^2 = -\dd t^2 + \cosh^2 t\, \dd\psi^2\,,
\ee 
where $\psi$ has a period of  $2\pi$.\footnote{As explained in \cite{Bobev:2018ugk}, for $p=1,2$ an analytic continuation must be performed whereby $\theta$ becomes timelike and $\psi$ spacelike such that the $\SU(1,1)$ symmetry is realized as the isometry of the hyperbolic plane.}  The ten-dimensional dilaton has the following form,
\be\label{10Ddilaton}
\e^{2\Phi} = g_s^2\e^{\f{p(7-p)}{6-p}\eta}~P~ Q^{\f{1-p}{2}}\,,
\ee
and the non-vanishing type II supergravity form fields are given by
\bea\label{eq:NSNSRRgenp}
B_2 &=&\e^{\f{p}{6-p}\eta}\f{YP}{g^2 X}\cos^3\theta~\vol_2\,,\\
C_{5-p} &=& \rmi\e^{-\f{p}{6-p}\eta}\f{YQ}{g_s g^{5-p}X}\sin^{4-p}\theta~\vol_{5-p}\,,\\
C_{7-p} &=& \f{\rmi}{g_sg^{7-p}}\left(\omega(\theta)+P\cos\theta~\sin^{6-p}\theta\right)\vol_2\wedge\vol_{5-p}\,.
\eea
Here $\vol_{5-p}$ and $\vol_2$ refer to the volume forms on $\dd \Omega_{5-p}^2$ and $\dd \widetilde{\Omega}_2^2$, respectively. The function $\omega(\theta)$ in \eqref{eq:NSNSRRgenp} is defined such that in the UV the exterior derivative of $C_{7-p}$ gives the volume form on the $(8-p)$--dimensional de Sitter space, namely
\be
\f{\dd}{\dd\theta}\left(\omega(\theta) + \cos\theta~\sin^{6-p}\theta\right) = (7-p)\cos^2\theta~\sin^{5-p}\theta\,.
\ee
%

\begin{figure}[H]
	\centering
	\includegraphics[scale=0.3]{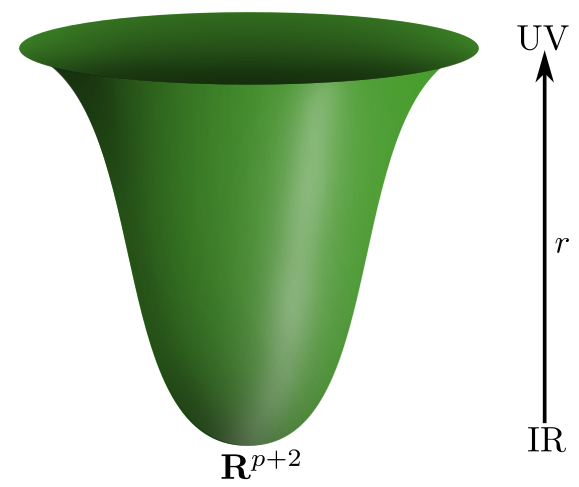}
	\caption{The regular geometries interpolate between flat Euclidean D$p$-branes in the UV and $\mathbf{R}^{p+2}$ in the IR.}
	\label{greencap}
\end{figure}

For a fixed value of $p$ the scalars $\eta(r)$, $X(r)$, and $Y(r)$ can be found by solving the BPS equations presented in Appendix~\ref{app:sugraspheres}. In the UV, i.e. for large values of $r$, the scalars $X$ and $Y$ take the values $X=1$ and $Y=0$ such that the solution asymptotically approaches the flat brane domain wall solution. In the IR region on the other hand, the solution is regular and the scalar fields approach a finite constant value. These IR values for the scalars can be found as the critical points of the superpotential \eqref{tildesuperpotential} and are given by:
\be
\begin{array}{lll}\label{eq:XYIR} X_\text{IR}= \f{p}{3}~,\quad& Y_\text{IR}=\pm\f{2(p-3)}{3}~,&\qquad\text{for $p<3~$}\,,\\
	X_\text{IR} = \f{p}{(6-p)(p-2)}~,\quad& Y_\text{IR} = \pm\f{2(p-3)}{(6-p)(p-2)}~,&\qquad\text{for $p>3~$}\,.\end{array}
\ee
Even though $X$ and $Y$ approach fixed values in the IR, the scalar $\eta$ can take any constant value $\eta_\text{IR}$.  A schematic form of the spherical brane solutions is depicted in Figure~\ref{greencap}.

An important ingredient in relating the supergravity results below to the ones found above using supersymmetric localization is the definition of 't Hooft coupling. In our conventions, the D$p$-brane tension and the Yang-Mills coupling constant are given in terms of the string coupling as
\be\label{gYMholo}
\mu_p = \frac{2\pi}{(2\pi\ell_s)^{p+1}}\,, \qquad  g_{\rm YM}^2 = \frac{(2\pi)^2 g_s}{(2\pi \ell_s)^4 \mu_p} = \frac{2\pi g_s}{(2\pi\ell_s)^{3-p}}\,.
\ee
The dimensionless holographic 't Hooft coupling, $\lambda_{\rm hol}$, is defined by
\be\label{philambdarel}
\lambda_{\rm hol}(E) = g_{\rm YM}^2 N E_{\rm hol}^{p-3}\,,
\ee 
where $N$ is the number of D$p$-branes and $g_{\rm YM}^2$ is defined in \eqref{gYMholo}. The quantity $E_{\rm hol}$ is a finite energy scale defined in an appropriate way through the supergravity solution. Since the supergravity backgrounds of interest here are not asymptotically locally AdS it is not straightforward to define this quantity. A reasonable choice is to define it as the inverse of the effective radius $R_{\rm eff}$ of the $(p+1)$-sphere $\dd\Omega_{p+1}^2$ in \eqref{10dmetricgeneral}, i.e.
\begin{equation}
R_{\rm eff} = Q^{-\frac{1}{4}}\e^{A+\frac{\eta}{2}}\,,
\end{equation}
and multiply it by the ten-dimensional dilaton $e^{\Phi}$  \eqref{10Ddilaton}. This definition amounts to the following result\footnote{We divide by a factor of $g_s$ since we have already included a factor of $g_s$ in the definition of $g_{\rm YM}^2$ in \eqref{gYMholo}.} 
\begin{equation}\label{eq:Eholdef}
E_{\rm hol}^{p-3} = R_{\rm eff}^{3-p} \frac{\e^{\Phi}}{g_s} = \e^{(3-p)A}\e^{\frac{9-p}{6-p}\eta} \frac{P^{1/2}}{Q^{1/2}}\,.
\end{equation}
This energy scale is finite in the UV limit $r\rightarrow\infty$ and thus we propose to identify the holographic 't Hooft coupling in \eqref{philambdarel} by evaluating \eqref{eq:Eholdef} in the UV where
\be
A \to \frac{(p-9)}{(6-p)(3-p)}\eta + \text{const}~.
\ee
The constant in this equation is fixed by regularity of the full supergravity background in the IR, it can therefore not be deduced directly by an UV analysis of the BPS equations.
Using that  $\lim_{r\rightarrow \infty} P(r) = \lim_{r\rightarrow \infty}Q(r) = 1$  we arrive at the following explicit result\footnote{An alternative way to obtain \eqref{lamhol} is to define the \emph{running} gauge coupling, as it appears in the probe action for D$p$-branes, by $g_{\rm YM}^2 = 2\pi \e^{\Phi}/(2\pi\ell_s)^{3-p}$. Then the energy scale is defined by $E = R_{\rm eff}^{-1}$. When these two expressions are inserted into \eqref{philambdarel} and evaluated at $r \to \infty$ we obtain \eqref{lamhol}.} 
\be\label{lamhol}
\lambda_\text{hol}\equiv  \frac{2\pi g_sN}{(2\pi\ell_s)^{3-p}}\e^{(3-p)A}\e^{\frac{9-p}{6-p}\eta}\bigg|_{r\to \infty}\,.
\ee	
We will sometimes express $\lambda_\text{hol}$ in terms of the supergravity gauge coupling $g$ using \eqref{eq:defg}. We note that the expression \eqref{lamhol}, which allows us to find a match between supergravity and field theory, does not agree with the one proposed in \cite{Kanitscheider:2008kd} for all values of $p$. 

\subsection{Holographic free energy}
\label{subsec:HoloFgen}

The holographic free energy of the spherical D$p$-brane solutions is given by the  on-shell action in $(p+2)$ dimensions. This action can be derived from the $(p+2)$-dimensional gauged supergravity, see \cite{Bobev:2018ugk}, and takes the form
\be\label{normalframe}
S = \f{1}{2\kappa_{p+2}^2}\int \dd^{p+2}x \sqrt{g}\left\{ R + \f{3p}{2(p-6)}|\dd \eta|^2 - 2{\cal K}_{\tau\tilde{\tau}}|\dd \tau|^2- V\right\}\,,
\ee
where the potential $V$ is given in Appendix \ref{app:sugraspheres} in terms of a superpotential $\mathcal{W}$. The $(p+2)$-dimensional Newton constant can be expressed as\footnote{This expression is derived in some detail in Appendix~\ref{app:kappa}.}
\be\label{NewtonsConstant}
\kappa_{p+2}^2 =  \f{(2\pi \ell_s)^8g_s^2}{8\pi}\f{\Gamma\left( \f{9-p}{2}\right)}{\pi^{\f{9-p}{2}}}g^{8-p}\,.
\ee 
Evaluating the action in \eqref{normalframe} on the spherical brane solutions leads to divergences arising from the UV region. Since for $p\neq 3$ the metric is not asymptotically locally AdS one cannot apply the standard technology of holographic renormalization to cancel these divergences systematically. As explained in \cite{Kanitscheider:2008kd,Kanitscheider:2009as} a useful approach to circumvent this impasse is to perform a conformal transformation of the metric to the so-called dual frame. This changes its UV asymptotics to the locally AdS form and for the solutions of interest here is achieved by the following rescaling
\be
g_{\mu\nu} = \e^{2a\eta}\tilde{g}_{\mu\nu} \,, \qquad\text{where}\quad a=\f{p-3}{6-p}\,.
\ee
Note that the case $p=6$ needs to be treated separately. For $p=3$ the background is asymptotically AdS$_5$ and no rescaling is needed. In terms of this transformed metric, the action takes the form
\begin{equation}\label{dualframe}
\begin{aligned}
S = \f{1}{2\kappa_{p+2}^2} \int \dd^{p+2}x \sqrt{\tilde{g}}\,\e^{p a \eta}\Big\{ &\tilde{R} + \big(\tfrac{3p}{2(p-6)}+ a^2p(p+1)\big)|\dd\eta^2| - 2{\cal K}_{\tau\tilde{\tau}}|\dd \tau|^2- \e^{2a\eta} V  \Big\}\, .
\end{aligned}
\end{equation}
In this frame the metric is asymptotically AdS and we can use the standard framework of holographic renormalization to obtain the holographic counterterm action. When transformed to the dual frame the Gibbons-Hawking boundary term is given by
\be\label{GHdual}
S_{\text{GH}} = \f{1}{\kappa_{p+2}^2}\int \dd^{p+1}x  \sqrt{\tilde{h}}\e^{ap\eta}(p+1)\left(A' - a\eta'\right)\, .
\ee
The remaining divergences should be cancelled by the standard curvature counterterms \cite{Emparan:1999pm}. However, as discussed in \cite{Kanitscheider:2009as}, the coefficients of these counterterms should be  changed with respect to the ones in \cite{Emparan:1999pm} and are determined by the constant $\sigma = \f{7-p}{5-p}$ . These infinite counterterms are built out of the induced boundary metric in the dual frame, $\tilde{h}_{\mu\nu}$ and are given by
\begin{multline}\label{curvct}
S_{\text{ct,curv}} = \frac{1}{\kappa_{p+2}^2} \int \dd^{p+1}x \sqrt{\tilde{h}}\e^{ap\eta}\Bigg[\f{2\sigma-1}{\sigma-1}g +  \f{1}{4g}R_{\tilde{h}} \\
+ \f{1}{16g^3}\f{\sigma-1}{\sigma-2} \bigg(R_{\tilde{h}ab}R_{\tilde{h}}^{ab}-\f{\sigma}{2(2\sigma-1)}R_{\tilde{h}}^2\bigg)  \Bigg]\,.
\end{multline}
The counterterms in the second line of \eqref{curvct} are only needed when $p\geq 4$. Note that this infinite counterterm analysis in the ``dual frame'' formalism is not applicable for $p=5$ and we will treat this case separately in Section~\ref{subsec:D5}.

 Apart from these curvature counterterms we typically need additional infinite counterterms coming from the scalar fields. For supersymmetric backgrounds we can take advantage of the Bogomol'nyi trick, see for example \cite{Freedman:2013ryh,Bobev:2013cja}, to construct these infinite counterterms. This amounts to adding the following counterterm built out of the superpotential of the gauged supergravity theory
\be\label{ctsuperpot}
S_\text{ct,superpot} = \f{1}{2\kappa_{p+2}^2}\int \dd^{p+1}x  \sqrt{\tilde{h}}\e^{(p+1)a\eta}\sqrt{\e^\mathcal{K}\mathcal{W}\overline{\mathcal{W}}}\Big|_{Y\rightarrow 0} \, .
\ee
This counterterm is precisely the one that appears when regularizing the free energy of supergravity backgrounds with flat space boundary. There might be additional counterterms appearing, such as conformal couplings of the scalars or terms depending on the scalar field $Y$, for more general solutions such as our spherical branes. The precise form of these extra infinite counterterms terms as well as any potential finite counterterms will be determined on a case-by-case basis in Section~\ref{sec:FandWLcases}. 

\subsection{Holographic Wilson loops}

Now let us demonstrate how to compute supersymmetric Wilson loop vacuum expectation values. The $\tfrac{1}{2}$-BPS Wilson loop captured by supersymmetric localization lies on the equator of the $(p+1)$-sphere and is invariant with respect to the localization supercharge if and only if it is aligned along the field theory scalar field $\phi_0$. This is realized by a fundamental string wrapping the equator of $S^d$ in the spherical brane solutions and embedded in a specific way in the internal space. To understand this in more detail we embed the internal space $I_{8-p}$ in $\mathbf{R}^{1,8-p}$,
\begin{equation}
\begin{aligned}
&X_I :\quad I_{8-p} \rightarrow \mathbf{R}^{1,8-p}:\\
&\{ \theta , t, \psi , \omega_i \} \mapsto \{ \cos\theta \sinh t,\,\cos\theta\cosh t\sin\psi,\,\cos\theta\cosh t\cos\psi,\,\sin\theta\, Y_A \}\,,
\end{aligned}
\end{equation}
where the $Y_A$ give the standard embedding of the $(5-p)$-sphere in $\mathbf{R}^{6-p}$. This embedding provides us with an explicit map from the internal space of our supergravity solutions to the field theory scalars appearing in the Lagrangian \eqref{Lss}, e.g. the scalars $\phi_I$ can be identified with $X_I$. Therefore, the BPS condition requires that the corresponding holographic Wilson loop lies at constant $\theta=0$ and $\cosh t= 0$. This implies that the holographic evaluation of the Wilson  loop VEV must be performed using the analytically continued fully Euclidean background. Indeed, this is how we obtained a finite Newton constant in \eqref{NewtonsConstant}.

\begin{figure}[H]
	\centering
	\includegraphics[scale=0.5]{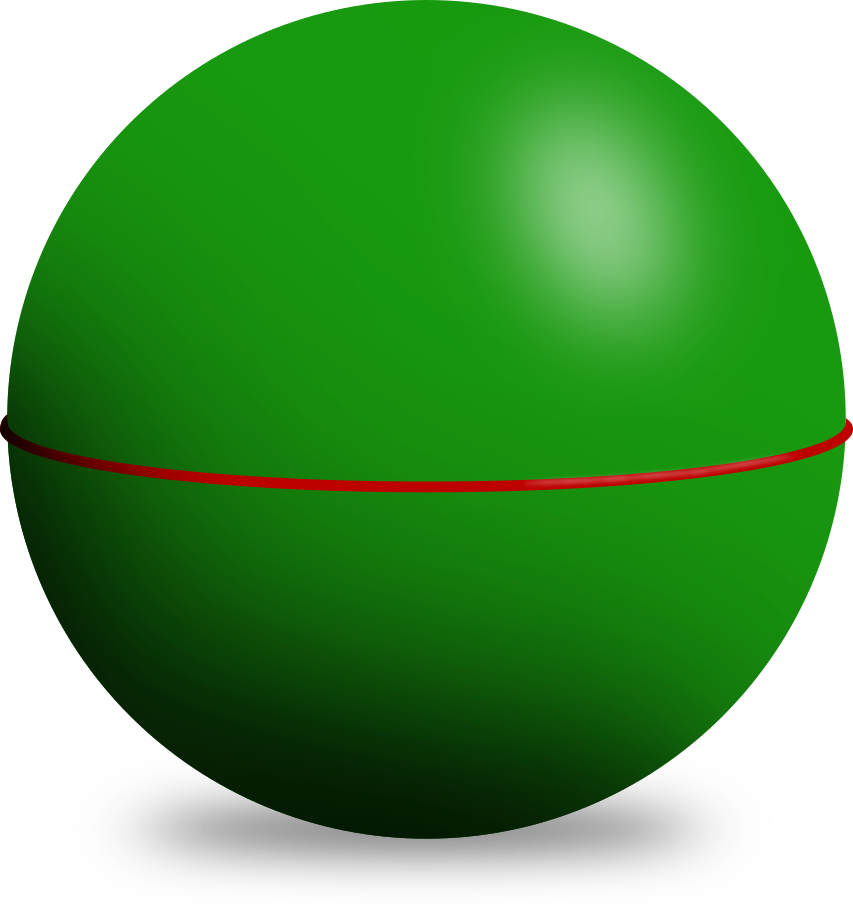}
	\caption{A string wrapping the equator of a $(p+1)$-sphere.}
	\label{Wilsondrawing}
\end{figure}

In the holographic context we are thus lead to study a probe fundamental string wrapping the equator of the spherical brane as in Figure~\ref{Wilsondrawing}. The expectation value of a Wilson line operator in the fundamental representation of the gauge group along a contour $\mathcal{C}$ can be calculated holographically by evaluating the regularized on-shell action of the probe string. More precisely,
\begin{equation}
\log \langle W(\mathcal{C})\rangle = - S^\text{Ren.}_{\text{string}}\,,
\end{equation}
where $S^\text{Ren.}_{\text{string}}$ is the renormalized on-shell action. The probe string is governed by the Nambu-Goto action,
\begin{equation}\label{eq:Sstringdef}
S_{\text{string}} = \f{1}{2\pi \ell_s^2}\int_\Sigma \dd^2\sigma \sqrt{\det P[G_{MN}]}-\f{1}{2\pi \ell_s^2}\int_\Sigma \det P[B_{2}] \, ,
\end{equation}
where $P[\dots]$ denotes the pull-back of the bulk fields onto the string worldsheet $\Sigma$ parametrized by $\sigma_1$ and $\sigma_2$ and $G_{MN}$ is the ten-dimensional string frame metric. In order to determine the Wilson loop expectation value we have to minimize the string action, regularize it and finally evaluate it on-shell. In order to do this, we parametrize the worldsheet by the coordinates $\sigma_1=r$ and $\sigma_2=\zeta \in [0,2\pi]$, use that translations along $\zeta$ are a symmetry of the ten-dimensional solution described in Section~\ref{subsec:sphericalbranes}, and assume that the induced fields depend only on $r$. Since $B_{2}$ has legs only along the internal de Sitter part of the geometry we conclude that $P[B_2] = 0$. The induced metric on the other hand takes the form
\begin{equation}
P[\dd s_{10}^2] = \frac{\e^\eta}{\sqrt{\mathstrut Q}}\left[\left( 1+ G_{mn}\frac{\partial\Theta^m}{\partial r}\frac{\partial\Theta^n}{\partial r} \right)\dd r^2 + \e^{2A}\dd\zeta^2\right]\,,
\end{equation}
where $G_{mn}$ is the metric on the internal space and the functions $\Theta^m(r)$ describe the profile of the string worldsheet in the internal directions. We can identify the functions $\Theta^m$ with the $8-p$ coordinates  $(\theta , t, \psi, \omega_i)$ with $i=1,\dots,5-p$. Minimizing the string action is equivalent to minimizing
\begin{equation}
\det P[G_{MN}] = \frac{\e^{2\eta+2A}}{Q}\left( 1+ G_{mn}\frac{\partial\Theta^m}{\partial r}\frac{\partial\Theta^n}{\partial r} \right)\,.
\end{equation}
Since we are performing the holographic computation for the ten-dimensional metric analytically continued to Euclidean signature,  
the internal metric $G_{mn}$ is positive definite. All terms in the parentheses above are therefore manifestly positive and thus can be minimized by setting each term to zero, i.e. by taking constant $\Theta^m$. To determine the exact position of the string in the internal space, i.e. the constant values of $\Theta^m$, we have to minimize the function
\begin{equation}
\begin{aligned}
\det P[G_{MN}]\big|_{\partial_r\Theta^m=0} =& \frac{\e^{2\eta+2A}}{Q} \\
=& \left\{\begin{array}{ll} \frac{\e^{2\eta+2A}}{X}\big(\sin^2\theta+X\cos^2\theta\big) & \text{for $p<3$}\,,\\
\frac{\e^{2\eta+2A}}{X} \big(X\cos^2\theta+(X^2-Y^2)\sin^2\theta\big) & \text{for $p>3$}\,.\end{array}\right.
\end{aligned}
\end{equation}
The extrema of these functions are at 
\begin{equation}
\theta = \f{n\pi}{2} \qquad \text{for } n\in\mathbf{Z}\, . 
\end{equation}
Since the range of $\theta$ is $[0,\pi)$ there are only two inequivalent extrema: $\theta = 0$ and $\theta=\pi/2$. However, as explained at the beginning of this section, only $\theta=0$ corresponds to a Wilson loop which is BPS with respect to the localizing supercharge.\footnote{See \cite{Bobev:2018hbq} for a similar analysis in the context of the four-dimensional $\mathcal{N}=2^*$ theory on $S^4$.}

We have thus arrived at the following probe string action \eqref{eq:Sstringdef}
\begin{equation}\label{WilsonGen}
S_{\text{string}} = \f{1}{\ell_s^2}\int \dd r \sqrt{\det P[G_{MN}]}= \f{1}{\ell_s^2}\int \dd r\e^{\eta+A}\,,
\end{equation}
where we have already performed the integral over the great circle. This on-shell string action diverges close to the UV boundary of the supergravity solution and we have to renormalize it using appropriate covariant counterterms built out of the ten-dimensional supergravity fields. This leads to the standard counterterm commonly used to regularize string on-shell actions \cite{Bobev:2018hbq,Drukker:1999zq}. In terms of the gauged supergravity fields, this counterterm takes the form
\be\label{WilsonCT}
S_{\text{string,ct}}=\f{1}{g \ell_s^2}\e^{A + \f{3}{6-p}\eta}\Big|_{r\to\infty}\,.
\ee
Note that in addition to cancelling the divergences of the on-shell string action, in some cases this counterterm contains a finite contribution which will prove to be crucial for our analysis. 

Before we discuss the various D$p$-branes in detail it is worthwhile to study how the Wilson line VEV scales with $N$ and $\lambda_\text{\rm hol}$. Using the scaling relation \eqref{philambdarel}, we find that 
\begin{equation}\label{eq:Sstringscaling}
\log\langle W\rangle \sim N^{0}\lambda_\text{hol}^{\frac{1}{(5-p)}}\, .
\end{equation}
This scaling exactly matches the expectations from supersymmetric localization. In addition the same scaling of the Wilson loop vacuum expectation value was found in a holographic finite temperature setting in \cite{Brandhuber:1998er}.

\section{Free energy and Wilson loop VEVs for spherical D$p$-branes}
\label{sec:FandWLcases}

After discussing the general framework for computing the free energy and Wilson loop expectation values, both from a supergravity and field theory point of view, we proceed with a case-by-case study of the different values of $p$, starting at $p=1$ and working our way up to $p=6$. For D5- and D6-branes some aspects of the general analysis above do not apply and we treat these two cases in some more detail. To avoid confusion, in this section we will denote the QFT 't~Hooft coupling in \eqref{tHooft} by $\lqft$ to explicitly distinguish it from the one used in supergravity denoted by $\lambda_\text{hol}$.

\subsection{D1-branes}\label{subsec:D1}

\subsubsection{Field theory}

In Section~\ref{sec:fieldtheory} we performed a general strong coupling analysis of the matrix model of \cite{Minahan:2015any} at large $N$.  
Strictly speaking, the  matrix model is only well defined for dimensions in the interval $3<d<6$.  To go below this interval let us first try returning to the general form of the kernel in \eqref{G16}.  If we set $d=2$ we find that the kernel takes the particularly simple form, 
\be\label{D1kernel}
G_{16}(\s)=\frac{4}{\s+\s^3}\,.
\ee
A matrix model with this kernel was previously analyzed in \cite{Kazakov:1998ji}  where the free energy was derived parametrically in terms of complete elliptic integrals.  However, in our case the central potential has a negative sign at $d=2$, which leads to many subtleties.  In particular a straightforward analytic continuation of the results in \cite{Kazakov:1998ji} gives a complex free energy in terms of $\lambda_\text{QFT}$.

Instead we propose to analytically continue  the dimension to $d=2$ in the expressions for the free energy and Wilson loop VEV in \eqref{FreeEn} and \eqref{WL}.   
Both the free energy and Wilson loop are expressible in terms of the eigenvalue endpoint, which  upon substituting $d=2$ into \eqref{bres} we find 
\be
b_2=\left(\frac{8 \lambda_{\rm QFT}}{\pi}\right)^{1/4}\,,
\ee
which is real and positive.
Having found $b_2$ we can read of the free energy from equation \eqref{FreeEn},
\be\label{eq:F2D1qft}
F_2=-\frac{2\pi}{3\lambda_{\rm QFT}}\left(b_2\right)^2N^2=-\frac{4(2\pi)^{1/2}}{3\lambda_{\rm QFT}^{1/2}} N^2\,.
\ee
Note that the free energy increases with increasing $\lambda_{\rm QFT}$. The Wilson loop VEV is obtained from \eqref{WLstr} by setting $b=b_2$  
\be\label{d2WL}
\log \langle W\rangle = {2\pi}b_2=2^{7/4}\pi^{3/4}\lambda_{\rm QFT}^{1/4}\,.
\ee
%

\subsubsection{Supergravity}

The supergravity solution for spherical D1-branes is most conveniently described using the scalar field $X$ as the radial variable. The full solution is then specified by
\begin{equation}\label{D1sol}
\begin{aligned}
Y^2(X) =&\, \f{(X + 1) (1 - X)^2}{X} \, , \\
\eta(X)  =&\, \eta_{\rm IR} + \f52 \log\f{1 - X}{2X}\,,\\
\e^A =&\, \frac{\sqrt{(1+X)^2-Y^2}}{g\e^{2\eta/5}}\frac{\sqrt{X}}{Y}\,,\\
X^\prime =& -\e^{2\eta/5}g\frac{\sqrt{X}(-2+2X^2+Y^2)}{\sqrt{(1+X)^2-Y^2}}\,,\\
\end{aligned}
\end{equation}
where the prime denotes a derivative with respect to the original radial coordinate $r$ and $X$ ranges from $1/3$ in the IR to $1$ in the UV. To compute the holographic free energy we evaluate the regularized supergravity action on the solution given above and subtract the counterterms \eqref{GHdual}, \eqref{curvct}, and \eqref{ctsuperpot}. In addition, due to the presence of the scalar $Y$ we have to subtract the following infinite counterterm
\be
S_\text{ct,inf} = -\f{1}{\kappa_3^2}\int \dd^2x \sqrt{\tilde{h}} \e^{-\frac{2}{5} \eta} \f{g}{4} Y^2\,.
\ee
Furthermore, there is a unique covariant finite counterterm that can be built out of the boundary metric and scalar fields which reads
\be
S_\text{ct,fin} = \f{1}{\kappa_3^2}\int \dd^2x \sqrt{\tilde{h}} \e^{-\frac{2}{5} \eta}\left(\f{c_x}{g} \tilde{R}\log X\right)\, .
\ee
Evaluating the holographic 't Hooft coupling \eqref{lamhol} in the UV leads to the following expression,
\begin{equation}\label{eq:lamholD1}
\lambda_{\rm hol} = \frac{1}{2^7 g^8 \ell_s^8\pi^3}\e^{4\eta_{\rm IR} /5}\,.
\end{equation}
Substituting this expression and subtracting all infinite and finite counterterms we arrive at the following result for the holographic free energy
\begin{equation}\label{eq:F2D1hol}
F^{\rm hol} = -\f{2(2\pi)^{1/2}N^2}{3\lambda_{\rm hol}^{1/2}}(3-4c_x) \, .
\end{equation}
We do not have a rigorous argument to fix the coefficient $c_x$ of the finite counterterm but we observe that if we set $c_x = 1/4$ the holographic result in \eqref{eq:F2D1hol} agrees with the field theory answer in \eqref{eq:F2D1qft} upon identifying $\lambda_{\rm hol}$ with $\lambda_{\rm QFT}$. It will be most interesting to fix $c_x$ by a first principle calculation. This can be presumably achieved by ensuring that the holographic renormalization procedure we employ is compatible with supersymmetry.

To compute the Wilson loop vacuum expectation value we start from the integral \eqref{WilsonGen}. For $p=1$ the on-shell probe string action becomes
\begin{equation}\label{D1gen}
S_{\text{string}} = \f{1}{\ell_s^2}\int_{1/3}^{1} \frac{\dd X}{X^\prime}\e^{\eta(X)+A(X)} = \f{\e^{\eta_{\rm IR}/5}}{\sqrt{2}g^2\ell_s^2}\int_{1/3}^{1}\frac{\dd X}{\sqrt{1-X^2}(1-X)}\,.
\end{equation}
This integral is divergent and we have to regularize it in the UV by introducing a cutoff at $X = 1-\epsilon$ and subsequently subtracting the counterterm \eqref{WilsonCT}
\be
S_{\text{string,ct}} =  \f{\e^{\eta_{\rm IR}/5}}{g^2 \ell_s^2}\f{1}{\sqrt{\epsilon}} + \mathcal{O}(\sqrt{\epsilon})\,,
\ee 
in order to obtain the renormalized on-shell action. Using the relation \eqref{eq:lamholD1} we find the following holographic result for the Wilson loop expectation value
\begin{equation}
\log \langle W^{\rm hol}\rangle  = 2^{7/4}\pi^{3/4}\lambda_{\rm hol}^{1/4}\,.
\end{equation}
This precisely agrees with the QFT result in \eqref{d2WL}.

\subsubsection{A comment on the Yang-Mills action}

We close this section with a comment.   In \cite{Benini:2012ui,Doroud:2012xw}  (see also  \cite{Closset:2014pda} for extensions of this analysis) it was shown that there is a Yang-Mills action for an $\mathcal{N}=(2,2)$ vector multiplet on $S^2$ that is $Q$-exact and hence the partition function is independent of the Yang-Mills coupling.  In terms of the conventions used here, the $\mathcal{N}=(2,2)$ vector multiplet contains the gauge fields $A_\mu$, the scalar fields $\phi_0$ and $\phi_3$, and the Dirac field $\Psi$ with the projections
\be
 \Gamma^{6789}\Psi=\Psi\,,\quad  \Gamma^{4567}\Psi=\Psi\,,
\ee 
which reduces $\Psi$ to four independent real components.  There is also one auxiliary field $K^1$. All other scalar and auxiliary fields are turned off.   If we restrict to four independent supersymmetry transformations where
\be
\Gamma^{6789}\eps=\eps\,,\quad \Gamma^{4567}\eps=\eps\,,
\ee
and set $\nu^1=\Gamma^{89}\eps$, the transformations on the fields in \eqref{susyos} reduce to
\bea\label{susyos2}
\delta_\eps\left( F_{12}-\frac{\phi_3}{\R}\right)&=&-\eps\Gamma_{12}\slashed{D}\Psi\nn\\
\delta_\eps\left(K^1-\frac{\phi_0}{\R}\right)&=&\eps\Gamma^{89}\slashed{D}\Psi\nn\\
\delta_\eps\Psi&=&\left(F_{12}-\frac{\phi_3}{\R}\right)\Gamma^{12}\eps+\left(K^1-\frac{\phi_0}{\R}\right)\Gamma^{89}\eps+D_\mu\phi_I\Gamma^{\mu I}\eps-\rmi[\phi_0,\phi_3]\Gamma^{03}\eps\nn\\
\delta_\eps \phi_{I}&=&\eps\Gamma_I\Psi\,.
\eea
It  is then straightforward to show that the flat-space Yang-Mills Lagrangian is invariant under the transformations in \eqref{susyos2} if
$F_{12}$ is replaced with $F_{12}-\frac{\phi_3}{\R}$ and $K^1$ is replaced with $K^1-\frac{\phi_0}{\R}$.  At the localization locus both terms are zero so the action is also zero.  

If we were to compare this Lagrangian to the one in \eqref{Lss} at $d=2$ and with the fields reduced as described above, then the Lagrangians differ by
\be\label{Lext}
-\frac{1}{2g_{\rm YM}^2}\Tr\left(\frac{2}{\R}F_{12}\,{\phi_3}-\frac{1}{\R^2}{\phi^3\phi_3}-\frac{3}{\R^2}\phi^0\phi_0-\frac{2}{\R}K^1\phi_0-\frac{1}{\R}\Psi\Lambda\Psi\right)\,.
\ee
One can show that \eqref{Lext} changes by a total derivative under the  supersymmetry transformations in \eqref{susyos2}.  Hence, both actions preserve $\mathcal{N}=(2,2)$ supersymmetry.  However, only the second action can be extended to 16 supersymmetries.  The extra term in \eqref{Lext} is not $Q$-exact so it will contribute a coupling dependent part to the partition function.

\subsection{D2-branes}
\label{subsec:D2}

\subsubsection{Field theory}

The matrix model analysis in this case is more subtle and one has to be careful when taking the different limits to obtain the kernel. If we set $d=3+\eps$ then we can approximate $G_{16}(\s)$ for $\eps \to 0$ as
\be\label{G3}
G_{16}(\s)=\frac{2\,\eps^2}{\eps^2\s+\s^3}+\frac{\pi\s(\coth(\pi\s)+\pi\s\csch^2(\pi\s))-2}{\s^3}\,\eps^2+{\rm O(\epsilon^3)}\,.
\ee
The first term in (\ref{G3}) comes from the $n=0$ term in (\ref{16susydet}) while the second term comes from all other values of $n$.  We can also see from (\ref{saddlept}) that $C_1\approx 4\pi^2\eps$ in this limit, which approaches zero because the super Yang-Mills action is $Q$-exact in three dimensions.    Aside from the first term, all other terms in (\ref{G3}) are nonsingular on the real line and of order $\eps^2$ or higher.  Hence they can be dropped in the saddle point equation in (\ref{saddlept}).  Therefore, in the large $N$ limit the saddle point equation reduces to the integral equation\footnote{After a rescaling the integral equation in \eqref{d3inteq} has the same form as in \cite{Kazakov:1998ji} and we could extract the  the free energy  by taking a limit of their results.}
\be\label{d3inteq}
\frac{4\pi^2\eps}{\lambda_{\rm QFT}}\s=2\pintd{-b}{b}\frac{\rho(\s')d\s'}{\s-\s'}-\int_{-b}^b\frac{\rho(\s')d\s'}{\s-\s'+\rmi\eps}-\int_{-b}^b\frac{\rho(\s')d\s'}{\s-\s'-\rmi\eps}+{\rm O}(\eps^2)\,.
\ee
Naively it looks like the right hand side of (\ref{d3inteq}) is even in $\eps$. However, because of the poles at $\s\pm \rmi\eps$ (\ref{d3inteq})  reduces to
\be
\frac{4\pi^2\eps}{\lambda_{\rm QFT}}\s=\pi \rmi\Big(\rho(\s+\rmi\eps)-\rho(\s-\rmi\eps)\Big)+{\rm O}(\eps^2)=-2\pi\eps\rho'(\s)+{\rm O}(\eps^2)\,.
\ee
Hence, to leading order in $\eps$ we have that $\rho(\s)=\frac{\pi}{\lambda_\text{QFT}}(b^2-\s^2)$.  The value of $b$ is fixed by setting
$\int_{-b}^b\rho(\s)d\s=1$, which gives
\be
b=b_3\equiv\left(\frac{3\lambda_{\rm QFT}}{4\pi}\right)^{1/3}\,.
\ee

The density $\rho(\s)$ and value for $b_3$ are precisely what one finds when analytically continuing \eqref{eigdens} and  \eqref{bres} to  $d=3$.
We can then use \eqref{FreeEn} and \eqref{WL} to find   the free energy and the expectation value of the BPS Wilson loop.  For the free energy we find
 \begin{equation}
\label{eq:FD2loc}
F_3=0\,,
\end{equation}
which is not surprising given the  $Q$-exactness of the SYM action  in three dimensions.  However, the Wilson loop is surprisingly nontrivial.  Here we find that 
\be\label{3exact}
\langle W\rangle=\frac{3}{\xi^3}\left(\xi\cosh\xi-\sinh\xi\right)\,,\qquad\xi=6^{1/3}\pi^{2/3}\lambda_{\rm QFT}^{1/3}\,.
\ee

To compare with supergravity we note that for for $\lambda_{\rm QFT}\gg1$ the logarithm of the Wilson loop VEV is approximately
\be\label{d3WL}
\log\langle W\rangle \approx 6^{1/3}\pi^{2/3}\lambda_{\rm QFT}^{1/3}\,.
\ee
We stress however that \eqref{3exact} is exact for any nonzero $\lambda_{\rm QFT}$.
If we expand \eqref{3exact} at small $\lambda_{\rm QFT}$ we find that
\be
\langle W\rangle=1+\frac{1}{10}(6\pi^2\lambda_{\rm QFT})^{2/3}+{\rm O}(\lambda_{\rm QFT}^{4/3})\,,
\ee
hence this result cannot be reproduced in perturbation theory.  Strictly speaking, the perturbative behavior is only found for $\lambda_{\rm QFT}<\eps^2$ where the matrix model approaches a Gaussian model.  In this sense,  $d=3$ MSYM is strongly coupled for any nonzero coupling.  

One can also see that the behavior of the Wilson loop VEV is essentially an infrared effect as the only relevant contribution to $G_{16}(\sigma)$ comes from the $n=0$ term in the partition function \eqref{16susydet}.  The numerator of this term is the Vandermonde determinant while the denominator is the uncanceled contribution of the constant spherical harmonics  about the localization locus \cite{Gorantis:2017vzz}.
%

\subsubsection{Supergravity}

The supergravity solution for spherical D$2$-branes is given by the following system of equations
\begin{equation}\label{eq:D2sugrasol}
\begin{aligned}
Y^2 =& \frac{1-X}{2X}\left((1-X)(1+2X)+\sqrt{(1-X)(1+3X)}\right)\,,\\
\e^\eta =& \e^{\eta_{\rm IR}}\frac{\sqrt{(1-X)\big(1+X+\sqrt{(1-X)(1+3X)}\big)}}{\sqrt{2}X}\,, \\
\e^A =& \frac{\e^{-\eta/4}}{g}\sqrt{\frac{X^3}{Y^2}-X}\,, \\
X^\prime =& -\e^{\eta/4}g\frac{\sqrt{X}(-2X+2X^2+Y^2)}{\sqrt{X^2-Y^2}}\,,\\
\end{aligned}
\end{equation}
Like for D$1$-branes we use $X$ as the radial variable which ranges from $2/3$ in the IR to $1$ in the UV.
In order to obtain the holographic free energy we proceed similarly to the previous case and subtract the counterterms \eqref{GHdual}, \eqref{curvct}, \eqref{ctsuperpot} and an additional infinite counterterm
\begin{equation}
S_{\text{ct,inf}} =-\f{1}{\kappa_4^2}\int \dd^3x \sqrt{\tilde{h}} \,\e^{-\frac{1}{2} \eta}\f{g}{4}\,Y^2 \, ,
\end{equation}
in order to obtain a finite free energy. In this case we do not find any finite counterterms. Evaluating the regularized on-shell action we find that the holographic free energy vanishes
\begin{equation}
F^{\rm hol} = 0\,.
\end{equation}
This agrees with the supersymmetric localization result in \eqref{eq:FD2loc}.

In order to compute the holographic Wilson loop expectation value we have to evaluate the following integral,
\begin{equation}
S_{\text{string}} = \f{1}{\ell_s^2}\int_{2/3}^{1} \frac{\dd X}{X^\prime}\e^{\eta(X)+A(X)}= \f{1}{g^2\ell_s^2}\int_{2/3}^{1} \dd  X \frac{\e^{\eta/2}(X^2-Y^2)}{Y(-2X+2X^2+Y^2)} \,.
\end{equation}
Using \eqref{eq:D2sugrasol} one can show that the integral reduces to 
\begin{equation}
S_{\text{string}} = \f{\e^{\eta_{\rm IR}/2}}{g^2\ell_s^2}\int_{2/3}^{1-\eps} \dd  X\frac{1}{\sqrt{(1-X)^3(1+3X)}}= \f{\e^{\eta_{\rm IR}/2}}{g^2\ell_s^2}\left(\frac{1}{\sqrt{\eps}}-\frac{3}{2}\right)\,,
\end{equation}
where we have introduced a cutoff $\epsilon \to 0$. To regulate the integral we need to subtract the counterterm \eqref{WilsonCT} given by
\be
S_{\text{string,ct}} = \f{\e^{\eta_{\rm IR}/2}}{g^2\ell_s^2}\left(\f{1}{\sqrt{\epsilon}} - \f12 + \mathcal{O}(\sqrt{\epsilon})\right)\,.
\ee 
Note that this counterterm contains a crucial finite piece needed to match the localization result. After substituting the explicit expression \eqref{lamhol} for $\lambda_{\rm hol}$, 
\begin{equation}
\lambda_{\rm hol} = -\frac{1}{6g^6\ell_s^6\pi^2}\e^{3\eta_{\rm IR} /2}\,,
\end{equation}
we find the following holographic result for the Wilson loop vacuum expectation value
\begin{equation}
\log \langle W^{\rm hol}\rangle  = \, -S_{\rm string}^\text{Ren.} = \, 6^{1/3}\pi^{{2}/{3}}\lambda_\text{hol}^{1/3}\,.
\end{equation}
This agrees with the field theory result \eqref{d3WL}.

\subsection{D3-branes}

The worldvolume theory on spherical D3-branes is simply the Euclidean $\mathcal{N}=4$ SYM theory on $S^4$. Since this is a conformal theory we can apply a conformal transformation to map $S^4$ to $\mathbf{R}^4$ and then analytically continue to Lorentzian signature.  Supergravity dual of the theory is the the well-known AdS$_5\times S^5$ background of type IIB supergravity. Both the QFT and supergravity evaluations of the free energy and Wilson loop vacuum expectation value are well-known results available in the literature. Here we briefly summarize how they can be obtained from our general formalism.

Setting $d=4$ in \eqref{bres} we find the eigenvalue endpoint 
\be
b_4=\frac{\sqrt{2\lqft}}{2\pi}\,,
\ee
which is the expected result from the Wigner distribution. To determine the free energy, we set $d=4+\eps$ and take the limit $\epsilon \to 0$ since there is a singularity in \eqref{FreeEn} at $d=4$. We find
\be
F_4= -\frac{2\pi^2N^2}{\lqft\,\eps}\left(\frac{\lqft}{2\pi^2}\right)^{1+\eps/2}+{\rm O}(\eps)= -\frac{N^2}{\eps}-\frac{N^2}{2}\log{\lqft}+{\rm O}(\eps)\,.
\ee
The divergent piece proportional to $\epsilon^{-1}$ is an overall constant that can be removed, leaving the well-known result for the Gaussian matrix model. The Wilson loop VEV can be found by inserting $b_4$ in \eqref{WL}
\begin{equation}\label{D3WL}
\log \langle W\rangle = \sqrt{2\lqft}\,.
\end{equation}
The free energy and the Wilson line VEV for $\mathcal{N}=4$ can also be computed holographically using standard results in the literature. An efficient way to obtain the end result on $S^4$ is to take the $m=0$ limit of the $\mathcal{N}=2^{*}$ calculations in \cite{Bobev:2013cja} and \cite{Bobev:2018hbq}.\footnote{Note that for $p=3$ our convention for $g_{\rm YM}^2$ as given in \eqref{gYMholo} differs by a factor $2$ from the convention used in \cite{Bobev:2013cja,Bobev:2018hbq}.}

\subsection{D4-branes}

\subsubsection{QFT}

Next we consider the case of spherical D4-branes which can be studied by setting $p=4$, or equivalently $d=5$, in the various general expressions above. From \eqref{bres} with $d=5$ we find that the eigenvalue endpoint is at
\be
b_5=\frac{\lqft}{4\pi^2}\,.
\ee
The free energy computed from \eqref{FreeEn} is then given by
\be\label{eq:F5qft}
F_5=- \frac{\lqft N^2}{12\pi}\,,
\ee
which agrees with the results in \cite{Kim:2012ava,Kallen:2012zn,Minahan:2013jwa}.

To compute the VEV of the BPS Wilson loop we need to plug the expression for $b_5$ in \eqref{WL} and take the large $\lambda$ limit to find 
\begin{equation}\label{eq:WLD4qft}
\log\langle W \rangle = \frac{\lqft}{2\pi}\,.
\end{equation}
%

\subsubsection{Supergravity}

The supergravity solution for spherical D4-branes is particularly simple as it is just a dimensional reduction of the AdS$_7\times S^4$ solution of eleven-dimensional supergravity. In this case the AdS$_7$ space has an $S^5\times S^1$ boundary. The spherical D$4$-brane solution is obtained by a reduction along $S^1$ leading to the following expression in our variables:
\begin{equation}\label{eq:D4soln}
\begin{aligned}
X =& 1\,,\\
Y =& \f12 \e^{2\eta_{\rm IR}-2\eta} \, ,\\
(\eta')^2 =& \f{g^2}{16}\e^{-5\eta}\left(\e^{4\eta}-\e^{4\eta_{IR}}\right)\,,\\
\e^A =& 2 \f{\e^{\eta/2}}{g}\sqrt{\e^{4\eta-4\eta_{IR}}-1}\,.
\end{aligned}
\end{equation}
Notice that it is convenient to use $\eta$ as the radial variable which runs from $\eta_\text{IR}$ in the IR to infinity in the UV.
To compute the holographic free energy we follow the, by now familiar, procedure of evaluating the on-shell action and subtracting the infinite counterterms \eqref{GHdual}, \eqref{curvct}, \eqref{ctsuperpot}. No other counterterms are required in order to regularize the action. However, we do find a number of covariant counterterms which give finite contribution to the on-shell action. These are given by\footnote{Two more finite counterterms can be written as a product of quadratic curvature invariant times $Y^2$, for an $S^5$ boundary, these are related to the last term in \eqref{D4finiteCT}.}
\be\label{D4finiteCT}
S_\text{ct,fin} = \f{1}{\kappa_6^2}\int \dd^5x \sqrt{\tilde{h}} \e^{2 \eta}\left(c_1 \left(\f{1}{g}\tilde R Y^2 -20g Y^4\right) + c_2 g Y^6 + \f{c_3}{g} \tilde{R} Y^4 + \f{c_4}{g^3} \tilde{R}^2 Y^2 \right)\, .
\ee
Although these counterterms look innocuous in six-dimensional gauged supergravity, from the perspective of the parent $\SO(5)$ gauged seven-dimensional supergravity, they are not gauge invariant. This is because the scalar field $Y$ arises as the component of one of the $\SO(5)$ gauge fields, $\mathcal{A}_\mu$, along the $S^1$ direction along which we reduce the seven-dimensional theory \cite{Bobev:2018ugk}. Therefore the $Y^2$ term in six dimensions corresponds to terms of the form ${\cal A}_\mu{\cal A}^\mu$. 
After adding all these contributions and substituting the 't Hooft coupling
\be
\lambda_{\rm hol} = \f{2\pi}{g^2\ell_s^2}\e^{2\eta_{\rm IR}}\,,
\ee 
the renormalized holographic free energy reads
\begin{equation}
F^{\rm hol} = - \frac{\lambda_{\rm hol}N^2}{96\pi}(10+80c_1+c_2+20c_3+400 c_4)\,.
\end{equation}
Similar to the discussion of the on-shell action for spherical D1-branes in Section~\ref{subsec:D1} we do not know how to fix the coefficients $c_{1,2,3,4}$ from a first principle calculation. However, we note that a convenient choice, namely 
\begin{equation}\label{eq:cD4}
c_1=c_2=c_4=0\,, \qquad c_3=-\f{1}{10}\,,
\end{equation}
makes the holographic result agree with the QFT calculation \eqref{eq:F5qft}. Gauge invariance of the seven-dimensional supergravity theory would indicate that all four finite counterterms should vanish, however in this case we reproduce the result of \cite{Kallen:2012zn} and do not find a match with the localization result. However, if we choose the counterterm coefficients as in \eqref{eq:cD4} we obtain an agreement with the supersymmetric localization calculation at the expense of breaking the gauge invariance of the supergravity counterterms. This predicament is reminiscent of the results in \cite{Genolini:2016sxe,Genolini:2016ecx} in the context of holographic renormalization for AdS$_{5}$ with an $S^3\times S^1$ boundary.

To evaluate the Wilson loop VEV we plug the solution \eqref{eq:D4soln} into the general expression \eqref{WilsonGen} using $\eta$ as a radial variable. We are then left with the following integral 
\begin{equation}
S_{\text{string}} = \f{1}{\ell_s^2}\int \dd r\e^{\eta+A} = \f{1}{\ell_s^2}\int \frac{\dd\eta}{\eta'}\e^{\eta+A} = \f{8}{g^2\ell_s^2}\int_{\eta_{\rm IR}}^{\infty} \dd\eta\, \e^{4\eta-2\eta_{\rm IR}}~.
\end{equation}
Evaluating the UV regulated integral and subtracting the counterterm in \eqref{WilsonCT} results in
\begin{equation}
\log \langle W^{\rm hol}\rangle = \frac{\lambda_{\rm hol}}{2\pi}\,,
\end{equation}
which matches with the localization result in \eqref{eq:WLD4qft}.

\subsection{D5-branes}
\label{subsec:D5}

\subsubsection{Field theory}

We next discuss MSYM on $S^6$ which was previously  investigated in \cite{Minahan:2017wkz}. This case is subtle because both \eqref{bres} and \eqref{FreeEn} have an essential singularity at $d=6$. 
To deal with this we set $d=6-\eps$, after which we find  
\be\label{F6}
F_6= -\frac{32\pi^4\eps N^2}{\lambda_b} \e^{-8/3-\gamma_E}\left(\frac{3\lambda_b}{8\pi^3\eps}\right)^{2/\eps}\,,
\ee
where $\gamma_E$ is the Euler-Mascheroni constant and $\lambda_b$ is bare 't Hooft coupling.
Hence  the free energy is negative and infinite for any value of $\lambda_b$ in the limit  $\eps\to0_+$.  
However, if we substitute $d=6-\eps$ directly into \eqref{saddlept} it takes the form to leading order in $\eps$
\bea
\label{eom:6d:nonrenorm}	
\frac{C_1}{\lambda_b}N\s_i=\left(\frac{6}{\eps}-6\gamma_E+4\right)N\s_i
-3\sum_{j\ne i}(\sigma_i-\sigma_j)\log(\s_i-\s_j)^2\,.
\eea
The first term on the right hand side can be absorbed into the 't Hooft coupling, hence we define the  renormalized coupling  
$\lambda_\text{QFT}$ in terms of $\lambda_b$ as
\be\label{lambdarenorm}
\frac{1}{\lambda_\text{QFT}}=\frac{1}{\lambda_b} + C_\lambda\,,
\ee
where the constant $C_\lambda$ is given by 
\be
C_\lambda=-\left(\frac{6}{\eps}-6\gamma_E+4\right)C_1^{-1}=-\frac{3}{8\pi^3}\left( \frac{1}{\eps}+\frac{1}{2}\log\left( 4\pi \right)-\frac{1}{2}\gamma_E-\frac{1}{3} \right)
\label{Clambda}
\ee
Notice that since the r.h.s. of (\ref{eom:6d:nonrenorm}) contains ${\eps}^{-1}$ 	it is crucial to  expand $C_1$  up to first order in $\eps$ to obtain $C_\lambda$ to order $\text{O}(\eps^0)$. 
Substituting $\lambda_b$ in terms of $\lambda_\text{QFT}$ into \eqref{F6} we find
\bea\label{F6r}
F_6&=&-12\pi \e^{-8/3-\gamma_E}N^2\left(1-\eps\left(\frac{8\pi^3}{3\lambda_\text{QFT}} - \frac{1}{3} - \frac{1}{2}\gamma_E +
\frac{1}{2}\log\left( 4\pi \right) \right)\right)^{2/\eps}
\nn\\
&=&-3 N^2\exp\left(-\frac{16\pi^3}{3\lambda_\text{QFT}}-2\right)\,.
\eea
The $-2$ in the argument of the exponent could be removed by a different scheme choice for $C_\lambda$.

A similar treatment can be applied to $b_6$.  Again  using \eqref{lambdarenorm} for $\lambda_\text{QFT}$ we find
\be\label{b6}
b_6=4\sqrt{\pi}\e^{-4/3-\gamma_E/2} \left(\frac{3\lambda_b}{8\pi^3\eps}\right)^{1/\eps}=2\exp\left(-\frac{8\pi^3}{3\lambda_\text{QFT}}-1\right)\,.
\ee
 This leads to the following expectation value for the BPS Wilson loop
\be\label{WL6}
\log\langle W\rangle\approx 4\pi \exp\left(-\frac{8\pi^{3} }{3\lambda_\text{QFT}}-1\right)\,.
\ee
The above results can also be directly obtained from the saddle-point equation (\ref{eom:6d:nonrenorm}) which we consider in detail in 
Appendix \ref{appendix:6d}. 
While   the prefactors of the exponentials in \eqref{F6r} and \eqref{WL6} are scheme dependent since they can be changed by a shift of
the renormalized coupling $\lambda_\text{QFT}$, we can take the following combination of  the free energy and the Wilson-loop VEV,
\be\label{6schemeindependent}
\f{F_6}{(\log\langle W\rangle)^2} = -\f{3 N^2}{16\pi^2}~,
\ee
which is scheme independent.

The form of (\ref{F6}) and (\ref{WL6}) is also suggestive.  We expect that the UV completion of 6D maximal super Yang-Mills is the (1,1) little string.  If we now write the free energy in terms of the little string tension
$T=\frac{2\pi^2}{g_\text{YM}^2}$, we get
\be
F_6\sim N^2 \exp\left(-\frac{16\pi^{3} }{3}\frac{T \R^2}{N}\right)\,.
\ee
In the large $\R$ limit  $S^6$ approaches flat space and $F_6$ falls off to zero, consistent
with the flat space free energy found in \cite{Cotrone:2007qa}.  The correction away from flat space
is suggestive of a non-perturbative contribution coming from the string world-sheet.  It would be interesting to explore this further.

Note that \eqref{b6} and the assumption that the eigenvalues are widely separated imply that $\lambda_\text{QFT}$ is small and negative. 
However,  (\ref{weakker}) and (\ref{saddlestrong}) show that near $d=6$ the crossover from the weak to the strong regime
happens when $|\s_{ij}|\sim \eps^{1/2}$.  The approximation is then valid if $b_6\gg \eps^{1/2}$ which corresponds to 
$\lambda_\text{QFT}\gg\frac{4\pi^{3 }}{3(-\log\eps)}$.  
Therefore, in the limit $\eps\to0$ the results in \eqref{F6} and \eqref{WL6} can be trusted for any positive 't~Hooft coupling.

\subsubsection{Supergravity}

As can already be seen from the localization computation, handling the divergences in this case is subtle. It is clear that the scaling relation in \eqref{eq:Sstringscaling} breaks down for $p=5$ and there are also special features of the supergravity solution which render the evaluation of the probe string action difficult. Additionally, the dual frame formalism of  \cite{Kanitscheider:2008kd} is not adapted to the case of five-branes. 

\begin{figure}
\centering
\begin{overpic}[width=0.75\textwidth]{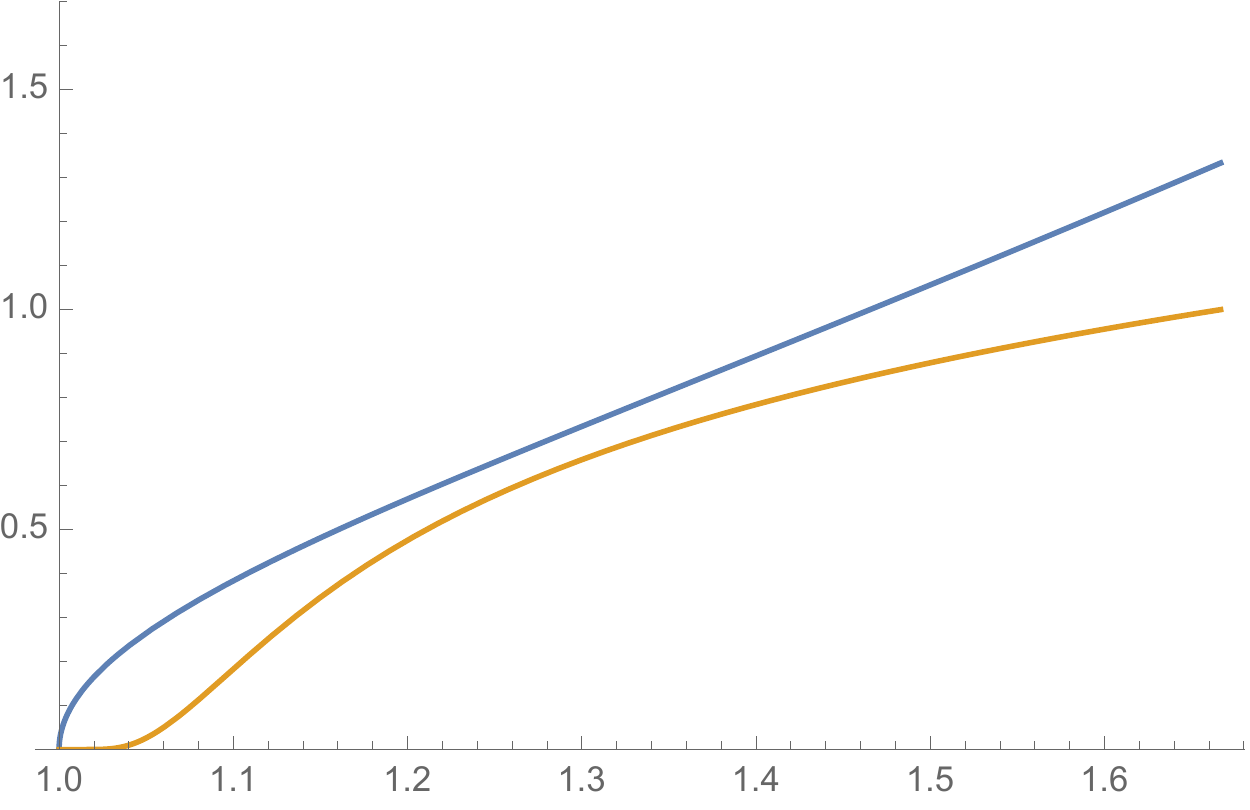}
\put(90,50) {$Y$}
\put(85,35) {$\e^{\eta_{\text{IR}}-\eta}$}
\put(100,0) {$X$}
\end{overpic}
\caption{\label{D5numerics}The supergravity solution for the scalar fields $Y$ and $\eta$ using $X$ as the radial coordinate.}
\end{figure}

The supergravity solution for spherical D5-branes can be obtained from the following system of equations
\begin{equation}\label{eq:D5soln}
	\begin{aligned}
	X^\prime =& \e^{-2\eta}\sqrt{X}g\f{2-8X+6X^2-3Y^2}{\sqrt{1-6X+9X^2-9Y^2}}\,,\\
	(Y^2)^\prime =& \e^{-2\eta}Y^2 g\f{1-16X+15X^2-9Y^2}{\sqrt{X(1-6X+9X^2-9Y^2)}}\,,\\
	\eta^\prime =& -\f{1}{10}\e^{-2\eta}g\f{\sqrt{1-6X+9X^2-9Y^2}}{\sqrt{X}}\,,\\
	\e^{2A} =& \f{\e^{4\eta}X((1-3X)^2-9Y^2)}{g^2 Y^2}\,.
	\end{aligned}
\end{equation}
We were not able to find an analytic solution to this system of equations. A numerical solution that interpolates between the IR at $(X,Y^2) = (4/3,16/9)$ and the UV at $(X,Y^2) = (1,0)$ is plotted in Figure~\ref{D5numerics}.
In order to extract holographic observables we must find asymptotic expansions for the supergravity fields. Unfortunately the BPS equations do not admit a simple UV expansion. Expressing $Y^2$ as a function of $X$ the asymptotic series is simultaneously an expansion in $(X-1)$ and $\e^{-1/(X-1)}$ of the general form
\be
Y^2 = P_0(X) + \e^{-1/(X-1)} P_1(X) + \e^{-2/(X-1)} P_2(X) + {\cal O}(\e^{-3/(X-1)})~.
\ee
where $P_i$ denotes a power series (possibly with negative powers) in $(X-1)$. The first two terms $P_0$ and $P_1$ can be resummed to yield
\begin{equation}\label{UVExact}
\begin{split}
Y^2 =& -\f{5X-6X^2+\sqrt{X(4-3X)}}{6}\\
&+ C \e^{-\f{1+\sqrt{X(4-3X)}}{X-1}}\f{2X+3X^2+3X\sqrt{X(4-3X)}}{\sqrt{X(4-3X)}}+ {\cal O}(\e^{-2/(X-1)})\,,
\end{split}
\end{equation}
where $C$ is a constant that must be carefully chosen so that the UV expansion matches onto the IR. Note that the first line in the above expansion is in fact an exact solution of the BPS equations, However, this solution does not reach the IR since one encounters a singularity at $X=4/3$. The corresponding UV expansion for $\eta$ takes the form
\be
\begin{split}
\eta =&\eta_\text{UV} +\f{1}{20}\left(\f{2+2\sqrt{X(4-3X)}}{X-1}+\log\f{2-X+\sqrt{X(4-3X)}}{4X} \right)\\
&-\f{6C\e^{-2/(X-1)}}{5(X-1)^2}+{\cal O}\left(\f{\e^{-2/(X-1)}}{X-1}\right)~.
\end{split}
\ee
In the IR we find an asymptotic series which follows the numerical solution to a very good approximation for a large part of the domain but deviates from the actual solution in the UV. This implies that the IR expansion will not be useful for extracting the holographic observables from the background. Instead we will employ the numerical solution. As we will explain, the linear behaviour of the ``dilaton'' $\eta$ in the UV will prevent us from performing a complete holographic renormalization as we did in
the 
 previous examples. 

First let us evalute the expression \eqref{lamhol} for $p=5$  to determine the relation of the supergravity parameters to the field theory data. Surprisingly we find that $\lambda_{\rm hol}$ does not depend on $\eta_\text{IR}$ at all. In fact we find
\be\label{D5lambdahol}
\lambda_\text{hol} = \lim_{X\to 1}\f{8\pi^3Y^2}{X((1-3X)^2 -9Y^2)} = \lim_{\epsilon\to0} \left(\f{8\pi^3\epsilon}{3} + {\cal O}(\epsilon^2)\right) = 0~,
\ee
where we use $X=1+\epsilon$ and $\epsilon \to 0^+$   
in  
the UV. Since this vanishes in the strict $\epsilon\to0$ limit we do not have a good definition of $\lambda_\text{hol}$ for D5-branes. We will therefore proceed with the computation of holographic observables and extract the $\lambda_\text{hol}$ by relating the localization and supergravity result for  
one of the 
observables, 
say the Wilson loop VEV. The relation can then be used to compare the supergravity result for 
the 
free energy with \eqref{F6r}.

Let us therefore evaluate the Wilson loop VEV for spherical D5-branes. In order to do so we can again use $X$ as a radial variable and evaluate the on-shell probe string action. Inserting the expressions \eqref{eq:D5soln} in the on-shell probe string action, we are left with the following integral 
\begin{equation}
S_{\text{string}} = \f{1}{\ell_s^2}\int \dd r\e^{\eta+A} = \f{1}{\ell_s^2}\int \frac{\dd X}{X'}\f{\e^{5\eta}}{Y}\f{(1-3X)^2-9Y^2}{2-8X+6X^2-3Y^2}~.
\end{equation}
Notice that the integrand depends exponentially on the dilaton $\eta$, in the UV this diverges as $5\eta = 1/\epsilon + {\cal O}(\log \epsilon)$. This implies that the integrand diverges in the UV  with a combination of polynomial and exponential powers in the cutoff $1/\epsilon$ 
\be
{\cal L}_\text{string} = \f{1}{g^2\ell_s^2} \left[\e^{1/\epsilon}\sqrt{3}\Big(\tfrac{1}{\epsilon} - 1 +{\cal O}(\epsilon)\Big) + {\cal O}(\e^{-1/\epsilon})\right]
\ee
where, as before, $X= 1+\epsilon$. Remarkably, the standard worldsheet counterterm, discussed around Eq. \eqref{WilsonCT} cancels the entire exponential divergence and leaves a finite on-shell action. Explicitly this counterterm has the form
\be
S_{\text{string,ct}} = \e^{5\eta}\f{\sqrt{X}\sqrt{(1-3X)^2-9Y^2}}{g^2\ell_s^2 Y}\bigg|_{X=1+\epsilon}~.
\ee
Once the action has been made finite in the UV we can evaluate it numerically using the numerical solution to the BPS equations. The accuracy of the numerical procedure is limited due to the fact that in the implementation of holographic renormalization we have to subtract large numbers. Nevertheless, we were able to show that with 1\% accuracy the following result holds
\be
\log \langle W^{\rm hol}\rangle = -S_{\text{string}}^\text{Ren.} \approx \f{1}{g^2\ell_s^2}\e^{5\eta_\text{IR}}~.
\ee
Comparing this expression with \eqref{WL6} suggests the relation
\be\label{D5newlambdahol}
\eta_\text{IR}= \f15\log(4\pi g^2\ell_s^2)-\frac{8\pi^{3} }{15\lambda_\text{QFT}}-\f15~.
\ee

Let us now return to the supergravity action with the aim to extract the holographic free energy. The UV analysis of the bulk supergravity action integrand has the following structure
\be\label{D5lagonshellschematic}
S_\text{on-shell} = \f{\pi^3}{5g^5\kappa_7^2}\left[\e^{2/\epsilon} \Big( \f{576}{\epsilon^5} + \f{1248}{\epsilon^4} + {\cal O}(\epsilon^{-3})\Big) + {\cal O}(\e^{1/\epsilon})\right] ~.
\ee
The polynomial divergence multiplying $\e^{2/\epsilon}$ can be cancelled by the standard covariant counterterms. However this still leaves seemingly infinitely many finite terms multiplying an exponential divergence.  In the case of five-branes, infinitely many counterterms are available due to the linear dilaton in the UV. It therefore seems that it is required to use infinitely many counterterms to eliminate the exponential divergence in \eqref{D5lagonshellschematic}. Indeed, we have not been able to find a finite set of counterterms that renders the action finite.\footnote{Such a finite set  of counterterms was shown to exist in a recent study of five-branes on some curved manifolds \cite{Aharony:2019zsx}.}
If we nevertheless assume that (finitely or inifinitely many) counterterms can be found that render the action finite, the form of the resulting expression can be deduced on general grounds. Since the bulk action is proportional to $\e^{10\eta}$ we expect
\be
S_\text{on-shell}^{\text{Ren.}} =- \f{6\pi^3\,{\cal I}}{g^5\kappa_7^2}\e^{10\eta_\text{IR}}~,
\ee
where ${\cal I}$ is an undetermined constant that we are not able to evaluate without a full knowledge of the counterterms. Using \eqref{D5newlambdahol} we find
\be
F^{\rm hol} = S_\text{on-shell}^{\text{Ren.}} = -3\,{\cal I} N^2 \exp\left[-\f{16\pi^3}{3\lambda_\text{QFT}}-2\right]~,
\ee
in a nice agreement with the field theory result \eqref{F6r}. As we argued above, the coefficients of the exponentials in \eqref{F6r} and \eqref{WL6} are dependent on the renormalization scheme but a scheme independent quantity can be found by combining the two as in \eqref{6schemeindependent}. We observe that the same combination in holography does not rely on the map \eqref{D5newlambdahol} and we find
\be
\f{S_\text{on-shell}^{\text{Ren.}}}{(S_{\text{string}}^\text{Ren.})^2} = - \f{3N^2{\cal I}}{16\pi^2}~,
\ee
which matches \eqref{6schemeindependent} if the constant ${\cal I}$ equals one.

\subsection{D6-branes}
\label{subsec:D6}

\subsubsection{Field theory}

We now turn to $d=7$ and start by rewriting the one-loop determinant in \eqref{16susydet} as
\be
Z_{\rm 1-loop} (\s)=\exp\left(\sum_{i<j}\sum_{n=1}^\infty2(n^2+1)\log\left(1+\frac{\s_{ij}^2}{n^2}\right)\right)\,.
\ee
To test the divergence we  expand the $\log$  at large $n$, showing that the log of the determinant behaves as
\be
\log Z_{\rm 1-loop} (\s)\sim \frac{1}{2}\sum_{i,j}\sum_n 2\s_{ij}^2\left(1+n^{-2}\right)-\s_{ij}^4\left(n^{-2}+n^{-4}\right)+\dots\,.
\ee
The sum over $n$ leads to a linear divergence for the $\s_{ij}^2$ term while the higher terms are finite.  The divergent piece can be rewritten as
\be \label{d7div}
2n_0\, N\sum_i\s_i^2\,,
\ee
where $n_0$ is a UV cutoff in $n$.   This divergence has the form of the action in \eqref{partfun} and can be absorbed
by shifting the coupling.  As for the D5-brane case, we can define a bare and a renormalized 't Hooft coupling through
the relation 
\be
\frac{1}{\lambda_\text{QFT}}=\frac{1}{\lambda_b}-\frac{n_0}{2\pi^4}\,.
\ee
%

\begin{figure}[!btp]
	\centering
	{\includegraphics[width=.9\linewidth]{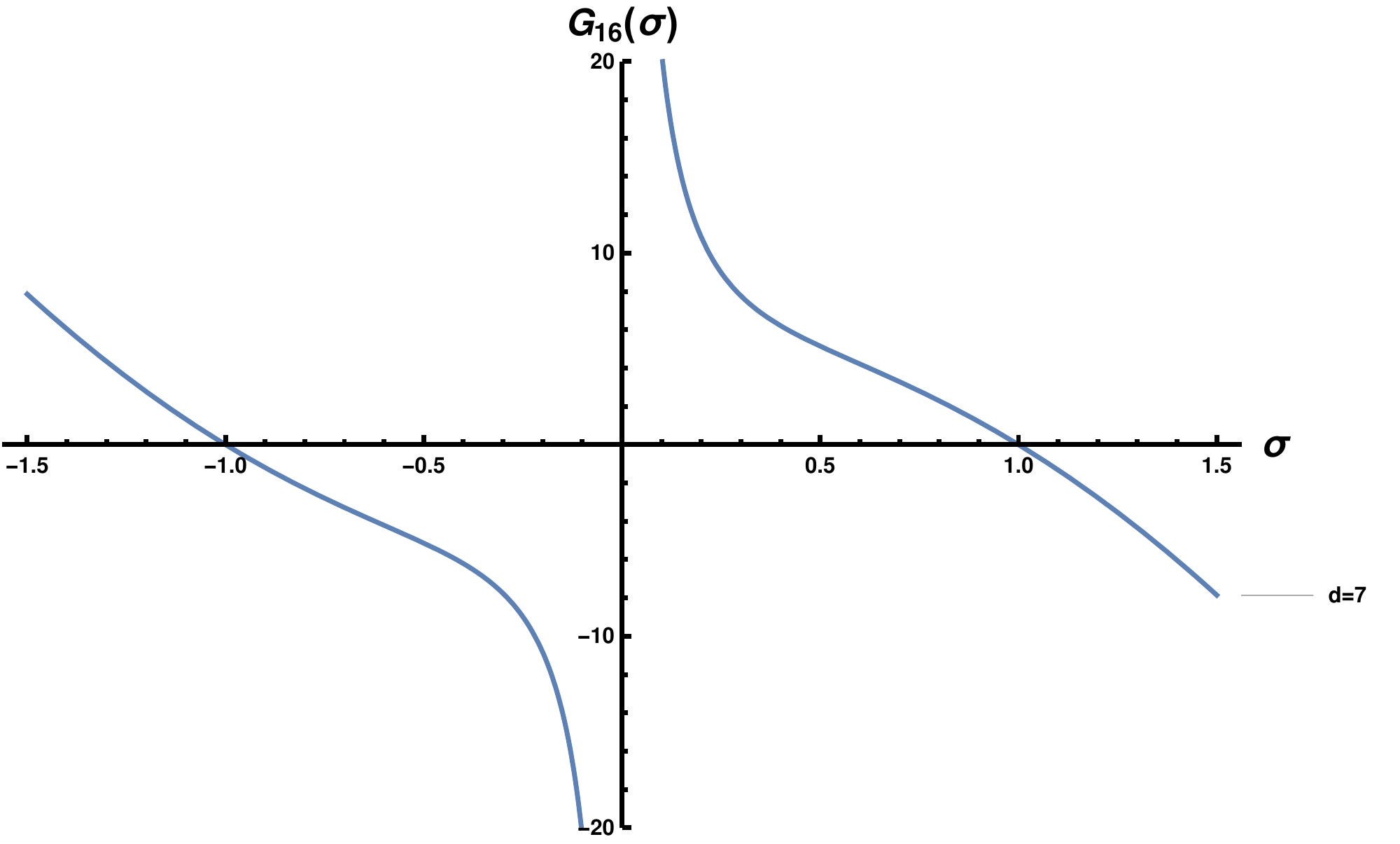}}
	\caption{The kernel $G_{16}(\s)$ for $d=7$.  At $|\s|=1$ the kernel crosses over from repulsive to attractive behavior.
	}
	\label{kernel7}
\end{figure}

The finite remainder from $Z_{\rm 1-loop} (\s)$ is what contributes to the analytic continuation of \eqref{G16} around the singularity
at $d=7$.  If we assume large separation between the eigenvalues then we can use \eqref{eom:strong} if we replace $\lambda$ with 
$\lambda_\text{QFT}$ in the lefthand side of the equation.  However,  $C_2$ in \eqref{C2def} is negative at $d=7$.  This is evident
in Figure~\ref{kernel7} which shows $G_{16}(\s)$ for $d=7$.  At short distance the kernel is repulsive, but becomes attractive for
$|\s|>1$.  Because of this negative sign, if we analytically continue  \eqref{bres} to $d=7$ we find that the eigenvalue endpoint is at 
\be\label{b7}
b_7=-\frac{ {2}\pi^3}{\lambda_\text{QFT}}\,.
\ee
The negative value for \eqref{b7} indicates that strictly speaking this is not a solution to \eqref{eom} assuming
the eigenvalue  distribution has the form in \eqref{densdelta}.  This is obvious since \eqref{eom} corresponds
to an attractive central potential and an everywhere attractive potential between the eigenvalues.  In this 
case the only solution  has all  eigenvalues at zero.  

\begin{figure}[H]
	\centering
	{\includegraphics[width=.6\linewidth]{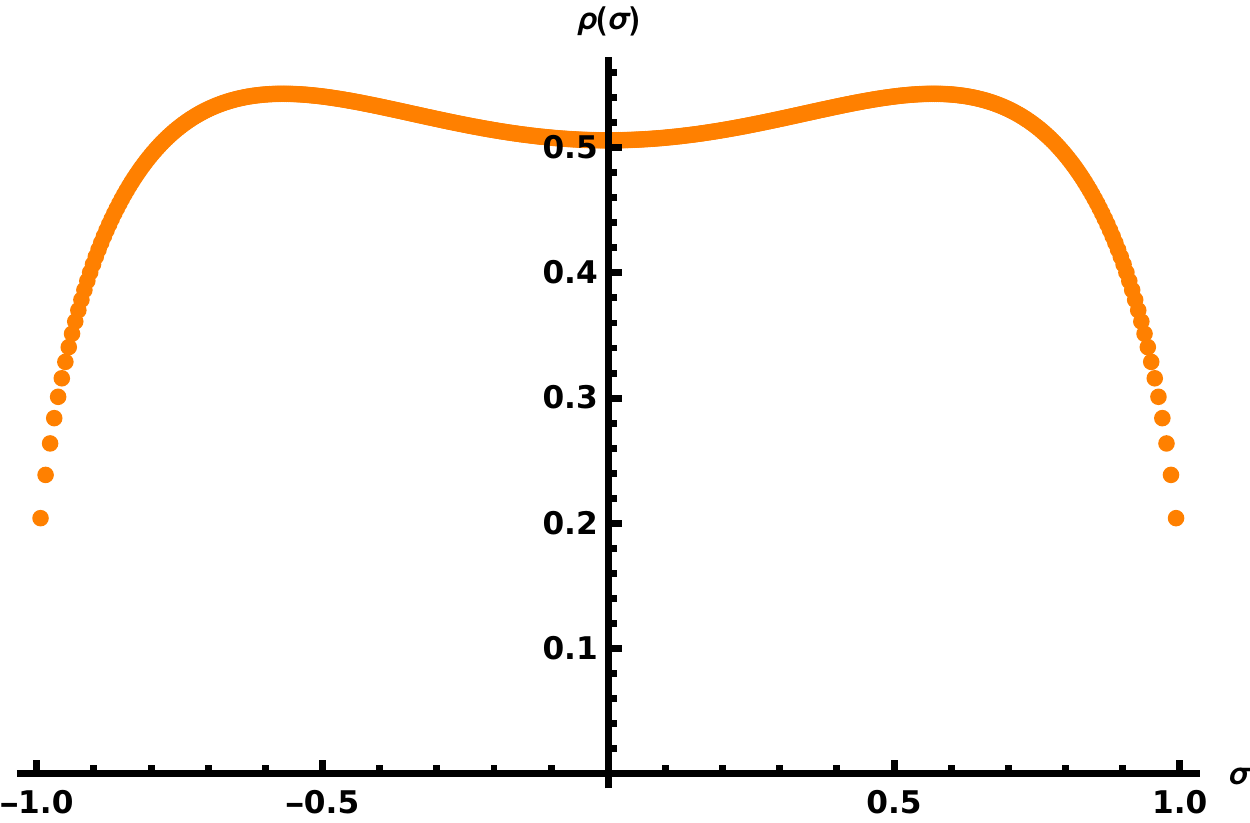}}
	\caption{Distribution of eigenvalues obtained from (\ref{saddlept}) numerically for $d=7$, $\lambda_\text{QFT}^{-1}=0$ and $N=501$.
	The eigenvalues are clearly bounded within a finite region.
	}
	\label{7d_inf_coup}
\end{figure}

To sort this out let us consider the  full $d=7$ kernel shown in Figure~\ref{kernel7},
\be\label{kerneld7}
G_{16}^{(7)}(\s\ms\s')=2\pi(1-(\s\ms\s')^2)\coth\pi(\s\ms\s')\,,
\ee
and take the strong  coupling limit so that the inverse renormalized coupling approaches $\lambda_\text{QFT}^{-1}\to0_+$.
While we cannot solve \eqref{saddlept} analytically in this limit, we can determine the eigenvalue distribution  numerically.
This is shown in Figure~\ref{7d_inf_coup} where we see that the short distance repulsion stabilizes the eigenvalues into a
bounded two hump distribution.  Hence the free energy  approaches a constant multiplied by $N^2$ in the strong coupling limit.  

Now suppose we continue $\lambda_\text{QFT}^{-1}$ through zero, such that  $\lambda_\text{QFT}<0$.   The central potential
is now repulsive and the eigenvalues are pushed farther away from the center, but are still stabilized by the attractive 
long-range force.  As we let $\lambda_\text{QFT}^{-1}$ become more and more negative the two humps in Figure~\ref{7d_inf_coup} get
pushed farther apart and we can then use the large separation approximation in \eqref{saddlestrong}.  In fact, in this approximation
the eigenvalue density becomes two delta functions, as is shown in appendix~\ref{appendix:7dalt}.   In appendix \ref{appendix:negcoup}
we numerically show  that the short range repulsion between the eigenvalues widens the delta functions to a width of order 1.

Since $\lambda_\text{QFT}<0$, $b_7$ in \eqref{b7}  is positive.  In order for the large separation assumption to be valid we require $b_7\gg1$, which happens 
when $\lambda_\text{QFT}^{-1}\ll-1$. Hence we are in a negative weak regime for the renormalized coupling, which is distinctly different from the usual positive
weak coupling regime.   Note  that while $\lambda_\text{QFT}^{-1}\ll-1$, $\lambda_b$ which appears in the original Lagrangian satisfies 
$\lambda_b^{-1}\gg 1$. If we now carry out the analytic continuation of \eqref{FreeEn} we find
\be\label{F7}
F_7=\frac{4\pi^4N^2}{3\lambda_\text{QFT}}\left(-\frac{\lambda_r}{2\pi^3}\right)^{-2}=\frac{16\pi^{10}N^2}{3\,\lambda_\text{QFT}^3}\,,
\ee
which diverges toward negative infinity as $\lambda_\text{QFT}^{-1}\to -\infty$.  Likewise, for the Wilson loop using \eqref{WL} we find that
\be\label{WL7}
\log\langle W\rangle = \log\cosh(2\pi b_7)\approx -\frac{4\pi^4}{\lambda_\text{QFT}}\,,
\ee
which  increases as $\lambda_\text{QFT}^{-1}\to -\infty$. Note that the $\cosh$ function is consistent with the delta function support at $d=7$.

Since the central potential is unbounded from below, the position of the eigenvalue center of mass is unstable. However, if the gauge group is
$\SU(N)$ and not $\U(N)$ then the eigenvalues satisfy the trace constraint 
$\sum_i \sigma_i=0$, which keeps the center of mass of the eigenvalues at the origin.
This suggests that the $\U(N)$ theory cannot be continued to negative $\lambda_\text{QFT}$.

Note further that the saddle point analysis is robust if $\lambda_\text{QFT}$ is small and negative, even when $N$ is finite.
As an example, consider an $\SU(2)$ gauge group. This has two eigenvalues $\s_1=-\s_2$,  and following the analysis in
appendix \ref{appendix:finiteN}  the saddle point gives the same free energy as \eqref{F7} with $N=2$.  From  \ref{appendix:finiteN}
we also see that the fluctuations to the free energy about the saddle point are
\be
\delta F=-\frac{4\pi^4}{\lambda_\text{QFT}}(\delta\s_1)^2\,,
\ee
hence the fluctuations are sharply suppressed and can be ignored if $-\lambda_\text{QFT}\ll 1$.

\subsubsection{Supergravity}

Spherical D6-branes do not fit in the general framework described in Section~\ref{sec:sugra}. In this case the $\SO(6-p)$ symmetry is trivial, hence the internal space is given simply by the two-dimensional de Sitter factor. Furthermore the eight-dimensional supergravity featured in the construction is particularly simple and contains only two scalar fields instead of the familiar three. For this reason we will analyze spherical D6-branes in several different ways in supergravity.

The spherical D6-brane solution is given by the following type IIA supergravity background, where we keep the radius $\mathcal{R}$ of the sphere arbitrary \cite{Bobev:2018ugk},
\begin{equation}\label{IIA7d}
\begin{aligned} 
\dd s_{10}^2 =& \f{{\cal R}^2\e^{2\Phi/3}}{g_s^{2/3}}\left(\f14\dd \rho^2 + \dd \Omega_7^2 + \f{1}{16}\sinh^2\rho ~\dd \widetilde\Omega_2^2\right)~,\\
H_3 =& \f{3}{g^2g_s^2}\e^{2\Phi}\dd\rho\wedge \vol_2~,\\
F_2 =& \f{\rmi}{g_sg}\vol_2~,\\
\e^{2\Phi} =& g_s^2\left(\f{g{\cal R}}{4}\sinh\rho\right)^3~.
\end{aligned}
\end{equation}
The radial coordinate $\rho$ takes values from $0$ to $\infty$. It is convenient at this point to define the new coordinate
\be
U\equiv\frac{2\pi^4{\mathcal R}^2\sinh^2\rho}{g_\text{YM}^2N}\,.
\ee
The equations in \eqref{IIA7d} then reduce to
\begin{equation}\label{IIA7dU}
\begin{split} 
\dd s_{10}^2 =&\ell_s^2\left(\left(\frac{g_{\rm YM}^2N}{2(2\pi)^4U}\right)^{1/2} \frac{\dd U^2}{1+\frac{g_\text{YM}^2NU}{2\pi^4 {\mathcal R}^2}}+ \left(\frac{2(2\pi)^4U}{g_{\rm YM}^2N}\right)^{1/2} \R^2\dd \Omega_7^2 + \left(\frac{g_{\rm YM}^2NU^3}{2(2\pi)^4}\right)^{1/2}  ~\dd \widetilde\Omega_2^2\right)\,,\\
H_3 =& \f{3\ell_s^2g_{\rm YM}^2NU}{(2\pi)^4{\mathcal R}}\frac{\dd U\wedge \vol_2}{\sqrt{1+\frac{g_{\rm YM}^2NU}{2\pi^4 {\mathcal R}^2}}}\,,\\
F_2 =& \f{\rmi N\ell_s}{2}\vol_2~,\\
\e^{2\Phi} =& \left(\frac{g_\text{YM}^2U^3}{2\pi^4N^3}\right)^{1/2}\,.
\end{split}
\end{equation}
These equations  reduce to the flat space supergravity solutions in \cite{Itzhaki:1998dd} when taking  $\mathcal{R}\to\infty$ while keeping  $U$ and $g_{\rm YM}^2N$ fixed.  The parameter $U$ can be thought of as the energy of a string stretched between a probe D6-brane and the $N$ D6-branes.  For small $U$ this is directly probing the weakly coupled 7D MSYM which has an effective coupling $g_\text{eff}^2=g_\text{YM}^2U^3$.   However,  in string units one sees that the curvature on the dS$_2$ is large for small $U$ so supergravity can not be trusted in this regime.

Following work of Susskind and Witten \cite{Susskind:1998dq}, Peet and Polchinski observed that $U$ is not actually the energy scale for a probe in supergravity \cite{Peet:1998wn}.  Instead, this is determined by the wave equation for a field in the bulk, say  a scalar $\psi$,  which is given by
\be
\left(-\frac{\partial^2}{\partial U^2}+\frac{k^2g_\text{YM}^2N}{2(2\pi)^4U}\right)U\psi=0\,,
\ee
where we have ignored the modes on dS$_2$.  From this we see that the energy scale for supergravity is
\be\label{PPsc}
E=\left(\frac{2(2\pi)^4}{g_\text{YM}^2NU}\right)^{1/2}\,.
\ee
In terms of this energy we have that effective coupling is 
\be
g_\text{eff}^2=g_\text{YM}^2\left(\frac{2(2\pi)^4}{g_\text{YM}^2NU}\right)^{3/2}\,,
\ee
which decreases with increasing $U$.  At the same time, the curvature on $dS_2$ is small if $U^3\gg \frac{2(2\pi)^4}{g_\text{YM}^2N}$, which corresponds to $g_\text{eff}^2N\ll 2(2\pi)^4$.  Hence it seems that the supergravity is dual to a weakly coupled gauge theory, but not the standard weakly coupled gauge theory since that is found at small $U$ where we cannot trust the supergravity.

Now let's assume that $\R$ is large but finite.  We then see from \eqref{IIA7dU} that we are in the flat brane regime when $E$ as defined in \eqref{PPsc} satisfies $E\gg \R^{-1}$.  This shows that an observer starts seeing the curvature of the branes when the energy scale is on the order of the inverse radius. Furthermore,  the radius of the $S^7$ should be small in string units, which requires that $E\ll \frac{2(2\pi)^4\R^3}{g_\text{YM}^2N}\R^{-1}$, hence we need weak coupling in order to trust the supergravity for distances significantly below the size of the sphere.  As $E$ approaches the sphere scale its dependence on $U$ starts to change, such that when $U\gg \frac{2\pi^4\R^2}{g_\text{YM}^2N}$, $E$ scales as $(\log U)^{-1}\sim \rho^{-1}$.

 As we keep increasing $U$ the string coupling eventually becomes large and we should uplift the solution to eleven-dimensional supergravity.  The uplifted metric and fields were given in \cite{Bobev:2018ugk}, where the solution takes the form of $\mathbf{H}^{2,2}/\mathbf{Z}_N\times S^7$. Explicitly, the eleven-dimensional metric is given by
\bea\label{metric}
\dd s_{11}^2&=&\frac{L^2}{4}\left(\dd s_4^2+4\dd\Omega_7^2\right)\,,\quad L= {\cal R}/g_s^{1/3}\nn\\   
\dd s_4^2&=&\dd\rho^2-\frac{\sinh^2\rho}{4}\left(\dd t^2-\cosh^2t\, \dd\psi^2+(N^{-1}\dd\omega-\sinh t \,\dd\psi)^2\right)\,.
\eea
This metric has two time directions, $t$ and $\omega$, which is to be expected since it describes the M-theory lift of a Euclidean brane. We refer to \cite{Bobev:2018ugk} for more details. The eleven-dimensional 4-form is given by
\begin{equation}\label{eq:G411d}
G_4 = \frac{6\,\rmi }{L}\, \vol_{\mathbf{H}^{2,2}}\,,
\end{equation}
where $\vol_{\mathbf{H}^{2,2}}$ is the volume form for the $\mathbf{H}^{2,2}/\mathbf{Z}_N$ metric.  The energy scale on the sphere maintains the  $\rho^{-1}$ fall off so that for large $\rho$ the only mode accessed is the constant one.
 Note that there is also a conical singularity at $\rho=0$ if $N>1$.  This singularity is what is left of the highly curved IIA theory at small $U$.

Let's now use the results from the previous section to propose a dual theory to the supergravity.  We saw using localization that there was a smooth transition between positive and negative $\lambda_r$.  We also saw that the ``strong coupling" behavior, that is having widely separated eigenvalues, occurs when $-\lambda_r\ll1$.   If we assume that $g_\text{YM}^2<0$ in the supergravity, then \eqref{eq:defg} and \eqref{gYMholo} imply that the metric and $\e^{2\Phi}$ in \eqref{IIA7d} have a negative sign.  To compensate for this we can send $\rho\to-\rho$ in which case we go back to the original signs for the metric and string coupling, while the $H_3$ field changes sign.  The eleven-dimensional supergravity metric in \eqref{metric} is unchanged but the   four-form field in \ \eqref{eq:G411d} changes by a sign under these transformations.  Hence, now everything looks almost the same as before, except any dictionaries we have between the supergravity and the gauge theory should have $g_\text{YM}^2$ replaced with $-g_\text{YM}^2$.  For example, the condition for small curvature on the dS$_2$ is now  $U^3\gg -\frac{2(2\pi)^4}{g_\text{YM}^2N}$, which translates to the relation $-g_\text{eff}^2N\ll 2(2\pi)^4$ for the effective coupling .

We are now ready to compute the free energy and Wilson loop VEV using supergravity.  One way to evaluate the free-energy of the spherical D6-brane is to use the eight-dimensional gauged supergravity originally used in \cite{Bobev:2018ugk} to construct the background. The eight-dimensional action is
\be
S = \f{1}{2\kappa_8^2}\int \star_8\bigg\{R-\f12(|\dd\beta|^2 + \e^{2\beta}|\dd \chi|^2 -\f{3g^2}{2}\e^{\beta}\bigg\}~,
\ee
where $\kappa_8^2$ is given by \eqref{NewtonsConstant}. The eight-dimensional BPS equations are written in terms of the metric
\be
\dd s_8^2 = \dd r^2 + {\cal R}^2\e^{2A}\dd\Omega_7^2~,
\ee
where the metric function $A$ is only a function of the radial variable $r$. The BPS equations read
\be
\begin{split}
\beta =&\, -6A~,\\
\chi' =&\, 6\rmi{\cal R}^{-1} \e^{-7A-2\beta}~,\\
(A')^2 =&\, {\cal R}^{-2}\e^{-2A} + \f{g^2\e^{\beta}}{16}~.
\end{split}
\ee
These equations can be easily solved by using the function $A$ as the radial variable. It can then be related to the coordinate $\rho$ appearing in \eqref{IIA7d} by the transformation
\be
\e^{2A} = \f{g{\cal R}}{4}\sinh\rho~.
\ee
Evaluating the action on-shell using the above expression for the eight-dimensional fields results in
\be
S = \frac{g{\cal R}^9}{2^{11}g_s^2 \ell_s^8 \pi^2}\int_{0}^{\infty}(1+7\cosh2\rho)\sinh\rho~\dd\rho~,
\ee
where we have included the Gibbons-Hawking term and performed the integral over the 7-sphere. This integral diverges as $\rho\to\infty$ which as we argued before is the IR of the geometry. The eight-dimensional metric is in fact completely regular there whereas the scalar $\beta$ diverges. This statement is of course dependent on the frame we use in supergravity. It is a lucky coincidence here that the Einstein frame metric is regular wheras the string frame or any other frame which is related to the metric above via a power of the scalar field $\e^\beta$ is singular. Subtracting divergences at $\rho \to \infty$ can therefore be done as before, by changing frame and introduce curvature counterterms such that the divergences cancel. We can also perform minimal subtraction; expand out the divergent terms and remove them by hand. In both cases the result is the same, the contribution of the IR is eliminated completely. The on-shell action is completely dominated by its contribution in the UV. Using the relations in \eqref{eq:defg} and \eqref{gYMholo} with $g_\text{YM}^2$ replaced with $-g_\text{YM}^2$ we find
\be\label{8DFE}
F^\text{hol} = \f{16 \pi^{10}N^2}{3\lambda_\text{hol}^3}~\,,
\ee
where
\be\label{D6lambdahol}
\lambda_{\rm hol} \equiv N g_\text{YM}^2 {\cal R}^{-3}=- \f{2^5\pi^4\ell_s^2}{g \mathcal{R}^3}\,,
\ee 
is defined as before but since there is no $\eta$-scalar in this case the equation \eqref{lamhol} is not directly applicable. The extra minus sign is to account for the negative Yang-Mills coupling. This result is 
in complete agreement with the localization result in \eqref{F7}.

We can also obtain the result in \eqref{8DFE} from the eleven-dimensional supergravity.
Before any Wick rotation the eleven-dimensional action is given by \cite{Hull:1998vg,Hull:1998ym}
\be\label{11dnewact}
S=\frac{1}{16\pi G_{11}}\int \dd^{11}x\sqrt{g^{(11)}}\left(R^{(11)}-\sfrac12|G_{4}|^2\right)\,,
\ee
where 
\be
|G_4|^2\equiv  \f{1}{4!}(G_4)_{\mu_1\dots\mu_4}(G_4)^{\mu_1\dots\mu_4}\,.
\ee
Substituting the solution \eqref{metric}-\eqref{eq:G411d} into \eqref{11dnewact} results in
\be
S=\frac{1}{16\pi G_{11}}\int \dd^{11}x\sqrt{g^{(11)}}\f{12}{L^2}\,.
\ee
In order to evaluate this on-shell action we need to Wick rotate one of the time directions as $t\to -\rmi\tau$ in \eqref{metric}. This changes the metric $\dd s_4^2$ to
\be\label{eq:ds2H31}
\dd s_4^2\to \dd\rho^2+\frac{\sinh^2\rho}{4}\left(\dd\tau^2+\cos^2\tau\, \dd\psi^2-(N^{-1}\dd\omega+\rmi\sin\tau \,\dd\psi)^2\right)\,,
\ee
Note that the M-theory circle parametrized by $\omega$ remains time-like.  Even though the metric is now complex,
its determinant remains real. The on-shell action then becomes\footnote{The contribution to the integral over $\rho$
in (\ref{SE1}) might not be trustable for $\rho\lesssim|\lambda_\text{QFT}|^{1/3}$.  However, if $|\lambda_\text{QFT}|\ll1$, then this
will lead to corrections of order $|\lambda_\text{QFT}|^{2/3}$ and the results in  (\ref{gF7n}) are  trustable to leading order.}
\be\label{SE1}
S_{\rm on-shell}=-\frac{1}{16\pi G_{11}}\int \dd^{11}x\sqrt{-g^{(11)}}\f{12}{L^2}=-\frac{1}{16\pi G_{11}}\f{\pi^6L^9}{2N}\int_0^{\rho_0}\dd\rho\sinh^3\rho\,,
\ee
where we have introduced a UV cutoff $\rho_0$ to regulate the volume of $\mathbf{H}_{3,1}$ in \eqref{eq:ds2H31}. As we take $\rho\gg1$ the on-shell action behaves as
\be
S_{\rm on-shell}=-\frac{1}{16\pi G_{11}}\f{\pi^6L^9}{2N}\left(\frac{1}{24} \e^{3\rho_0}- \frac{3}{8} \e^{\rho_0} +\frac{2}{3}+\mathcal{O}(\e^{-\rho_0})\right)\,.
\ee
The divergent contributions in this expression should then be removed to obtain a finite action. Using $G_{11}=16\pi^7\ell_{s}^9$ and the modified AdS/CFT dictionary to account for the negative coupling, $(2\pi\ell_{s})^3g_s=-\frac{g_{\rm YM}^2}{2\pi}$, we find\footnote{Note that in \cite{Bobev:2018ugk} the regularized on-shell action was computed using a four-dimensional effective supergravity approach leading to a result which differs by a factor of 2 from \eqref{gF7n}. The eight-dimensional and eleven-dimensional approaches we use here is better justified and should be employed instead.}
\be\label{gF7n}
S_{\rm on-shell}^\text{Ren.}=\frac{16\pi^{10}}{3\lambda_{\rm hol}^3}N^2\,.
\ee
This again agrees nicely with the free energy in \eqref{F7}.

The BPS Wilson loop can be computed using the IIA solution in (\ref{IIA7d}). The on-shell string action in this case is given in terms of the eight-dimensional metric function
\begin{equation}
S = \f{1}{\ell_s^2}\int \mathcal{R}\e^A\,\dd r \,.
\end{equation}
Changing coordinates to the radial coordinate $\rho$ as above we find
\begin{equation}
S_{\text{string}} = \f{g \mathcal{R}^3}{8\ell_s^2}\int_{0}^{\infty}\dd\rho\sinh\rho\,,
\end{equation}
Just like the on-shell action, this integral diverges in the IR and can be regularized by adding a simple counterterm analogous to \eqref{WilsonCT}. This counterterm implements minimal subtraction resulting in the following expression for the Wilson line expectation value
\begin{equation}\label{gWL7}
\log\langle W^{\rm hol}\rangle = -\f{4\pi^4}{\lambda_{\rm hol}}\,,
\end{equation}
where we have again flipped the sign of $g_\text{YM}^2$ in the dictionary.
This precisely agrees with the localization result in \eqref{WL7}. 

An alternative way to compute the Wilson loop VEV is to evaluate the on-shell action of an appropriately embedded M2-brane in the eleven-dimensional solution \eqref{metric}. The M2-brane wraps the equator of $S^7$ and extends along $\rho$ and the M-theory circle $\omega$. In particular the brane is fixed along $t$ and $\psi$ since it should be constant along the field theory scalar $\phi_0$. The holographic dual to the Wilson loop VEV is then given by
\begin{equation}
	\log\langle W^{\rm hol}\rangle = -S_{M2}^\text{Ren.}\,,
\end{equation}
where the probe M2-brane on-shell action is given by
\be
	S_\text{M2} = \mu_2\int \dd^3 \sigma \sqrt{\det P[G_{MN}]}\,.
\ee 
$P[ G_{MN}]$ denotes the pullback of the determinant of the eleven-dimensional metric to the M2-brane worldvolume and the brane tension is given by $\mu_2=\f{2\pi}{(2\pi\ell_s)^2}$. Evaluating this action on our solution gives the following diverging result
\begin{equation}
	S_\text{M2} = \mu_2\f{L^3}{4}\f{2\pi}{N}(2\pi)\int_{0}^{\infty}\dd\rho \sinh\rho = \f{g L^3}{8\ell_s^2}(\cosh\rho_0-1)\,.
\end{equation} 
where in the second step a cut-off $\rho_0$ was introduced to regulate the divergence. By adding a simple counterterm
\begin{equation}
	S_\text{W,ct} =- \f{g L^3}{8\ell_s^2}\sinh\rho_0\,,
\end{equation}
very similar in spirit to the counterterm \eqref{WilsonCT}, we obtain the following renormalized on-shell action
\begin{equation}\label{eq:SWLrenD6}
	S^\text{Ren.}_{\text{M2}} =-\f{g L^3}{8 \ell_s^2}\,.
\end{equation}
Inserting the expression for $\lambda_{\rm hol}$ in this equation with a sign change in the dictionary results in the following expression for the holographic Wilson loop VEV
\begin{equation}\label{gWS7}
	\log \langle W^{\rm hol}\rangle = -\f{4\pi^4}{\lambda_{\rm hol}}\,.
\end{equation}
This agrees nicely with the type IIA calculation in \eqref{gWL7} and the localization result in \eqref{WL7}.

In \cite{Itzhaki:1998dd} it was noted that the supergravity solution for D6-branes could be trusted even for small $N$.  This is consistent with our results here.   As we showed in the last section, the form of the free energy holds for small $N$, at least if $N$ is even.  In the classical supergravity we find the same free energy as a function of $N$ so this appears to align well with the claim in \cite{Itzhaki:1998dd}.  There is  a subtlety however for odd $N$.  As follows from \eqref{Ffinite}, the localization result for the free energy comes with an extra factor of $\frac{N(N-1)}{(N-1/2)^2}$.  This arises because one eigenvalue has to be placed at the origin in the solution to the saddle point equation.  Hence, it is essentially a quantization condition that the supergravity does not directly see.  In the $N=1$ case the free energy is zero for the gauge theory, not surprisingly since the gauge group is $\SU(1)$ which is trivial.  The supergravity does not look trivial although the eleven-dimensional uplift is now smooth at the origin.  It would be interesting to understand this point better.

\section{Discussion}
\label{sec:discussion}

In this work we showed how to compute the partition function and the expectation value of a BPS Wilson loop for $\SU(N)$ maximal SYM on $S^d$ in the limit of large $N$ and large 't Hooft coupling for $2\leq d \leq 7$. We approached this problem using supersymmetric localization in the QFT as well holography using the spherical brane supergravity solutions of \cite{Bobev:2018ugk}. Both calculations involve non-trivial elements due to the peculiarities of the localization matrix model for certain values of $d$ and the fact that the supergravity solutions are not asymptotically AdS. It will thus be interesting to extend and generalize our work in several directions which we discuss  below.

Studying the generalization of our construction for SYM theories with less supersymmetry is of clear interest. Both pure and matter coupled SYM theories on $S^d$ with 8 supercharges exist in $2\leq d\leq6$ and it is possible to study them in the large $N$ limit using supersymmetric localization. Constructing the corresponding supergravity solutions is not straightforward since it is not clear which classes of such SYM theories have weakly coupled supergravity dual. The analogous question for SYM with 4 supercharges on $S^d$ with $d\leq4$ is also interesting but is perhaps even harder to analyze, both in quantum field theory and in supergravity. We do not know how to extend our construction to $d>7$ but it will certainly be interesting to study this. See \cite{Prins:2018hjc} for some recent work on curved D7-branes and \cite{Minahan:2018kme,Naseer:2018cpj} for a QFT analysis that may be relevant to this question. We have focused on MSYM theories on the round sphere in this paper. It is possible to place supersymmetric gauge theories on other curved manifolds, for example on squashed spheres, at the price of breaking part of the supersymmetry. The generalization of our analysis to these more general setups will be interesting to pursue.

In the analysis of the matrix model results we have focused on planar MSYM in the limit of large 't Hooft coupling $\lambda$. It will be very interesting to extend this analysis to finite values of $\lambda$ while remaining in the large $N$ limit. This will allow us to understand whether there are any interesting phase transitions as a function of $\lambda$ akin to the ones observed for $\mathcal{N}=2^{*}$ in \cite{Russo:2013qaa,Russo:2019lgq}.  For MSYM on $S^3$ our results appear to be exact in $\lambda$. While the free energy of this theory vanishes the Wilson loop VEV is non-trivial and does not have a form suggesting a non-trivial phase transition. It will be desirable to understand this result from a a perturbative analysis in the weakly coupled planar MSYM theory. Alternatively one could attempt to study $1/\lambda$ corrections to the on-shell action of the probe-string in the supergravity solution.

In the holographic analysis of the spherical brane solutions we successfully employed the holographic renormalization procedure in the context of asymptotically non-AdS space-times. It is desirable to put this procedure on a more solid footing and to address the subtle question on how to fix the coefficients of the finite counterterms in the on-shell actions we have encountered for D2- and D4-branes.

Finally, we would like to stress that for D5/NS5-branes, both in the matrix model and the supergravity solution, we have encountered some intriguing UV divergences which we managed to regularize in a seemingly consistent way. It will be very interesting to understand whether these calculations can teach us something about the structure of the $(1,1)$ little string theory.

\bigskip
\bigskip
\leftline{\bf Acknowledgements}
\smallskip
\noindent We are grateful to Davide Cassani, Guido Festuccia, Elias Kiritsis, Gregory Korchemsky, Jesse van Muiden, Nikita Nekrasov, Silviu Pufu and Konstantin Zarembo for interesting discussions. The work of NB and 
PB is supported in part by the Odysseus grant G0F9516N from the FWO. FFG is a Postdoctoral Fellow of the Research Foundation - Flanders (FWO). Furthermore, this work is supported in part by the KU Leuven C1 grant ZKD1118 C16/16/005. The research of JAM is supported in part by Vetenskapsr{\aa}det under grant \#2016-03503 and by the Knut and Alice Wallenberg Foundation under grant Dnr KAW 2015.0083. JAM thanks the Center for Theoretical Physics at MIT and Nordita for kind hospitality during the course of this work. AN is supported by the Israel Science Foundation under grant No.~2289/18, and by the I-CORE Program of the Planning and Budgeting Committee. 
AN thanks Uppsala University for kind hospitality during the course of this work.

\appendix 
\section{Useful integrals}
\label{useful}

The following integrals are useful for the calculations in Section~\ref{sec:fieldtheory} 
\be\label{intstr}
\pintd{-b}{b}\frac{d\s'|\s-\s'|^\al\sign(\s-\s')}{(b^2-{\s'}^2)^{\al/2}}={\pi\al\s}\csc\left(\tfrac{\pi\al}{2}\right)\,,
\ee
and
\be\label{rhoint}
\int_{-b}^b\frac{d\s'}{(b^2-{\s'}^2)^{\al/2}}=\frac{b^{1-\al}\pi^{1/2}\Gamma(\tfrac{2-\alpha}{2})}{\Gamma(\tfrac{3-\alpha}{2})}\,,
\ee
where we have defined  $\al\equiv d-5$. Note that the result in \eqref{intstr} is independent of $b$. The result in \eqref{intstr} can be understood by splitting the integral into two parts,
\be\label{intstr2}
\int_{-b}^{\s}\frac{d\s'(\s-\s')^\al}{(b^2-{\s'}^2)^{\al/2}}-\int_{\s}^{b}\frac{d\s'(\s'-\s)^\al}{(b^2-{\s'}^2)^{\al/2}}\,.
\ee
Both integrals in \eqref{intstr2} are discontinuous as $\s$ crosses the branch cuts between $-\infty<\s<-b$ or $b<\s<\infty$.  However, it is straightforward to show that the discontinuities cancel between the two integrals and so the sum must be a holomorphic function of $\s$ in the complex plane.  By taking  $\s$ to a large imaginary value in \eqref{intstr2} one can see that the combined integrals have a leading linear behavior in $\s$ with the coefficient in \eqref{intstr}, while the constant piece is zero because the integral in \eqref{intstr} is clearly an odd function of $\s$.

It proves useful to define the following function
\be\label{fdef}
f(\s)\equiv\frac{\Gamma(\tfrac{3-\alpha}{2})}{\pi^{1/2}\Gamma(\tfrac{2-\alpha}{2})}\frac{\s}{b}\,{}_2F_1\left(\frac{1}{2},\frac{\al}{2};\frac{3}{2};\frac{\s^2}{b^2}\right)\,.
\ee
One can show that $f'(\s)=\rho(\s)$ where $\rho(\s)$ is defined in \eqref{eigdens}, and that $f(b)=1/2$.  Note that since $\rho(\s)$ is an even function of $\s$, $f(\s)$ is an odd function.

Finally, we present two integrals which are useful for the calculation of the free energy
\be\label{I_1def}
I_1\equiv\int_{-b}^bd\s \rho(\s)\s^2=\frac{b^2}{8-d}\,,
\ee
and
\be\label{I_2def}
I_2\equiv\int_{-b}^bd\s \rho(\s)(b-\s)^{d-4}=2b^{d-4}\pi^{-1/2}\Gamma(\tfrac{d-1}{2})\Gamma(\tfrac{8-d}{2})\,.
\ee
%

\section{Gauged supergravity construction}
\label{app:sugraspheres}

Here we summarize the results in \cite{Bobev:2018ugk} on how to obtain the spherical brane solutions of interest from maximal gauged supergravity in $p+2$ dimensions. The Lorentzian supergravity theory is constructed by a reduction of type II supergravity on $S^{8-p}$. This theory has to then be analytically continued to Euclidean signature and further truncated to its $\SU(1,1)\times \SO(6-p)$ invariant sector in order to capture the R-symmetry of the dual maximal SYM theory on $S^{p+1}$. The spherical brane solutions preserve the $\SO(p+2)$ isometry of the sphere which the branes are wrapping. This implies that the solutions can be constructed by restricting only to the metric, one real scalar fields $\eta$, and one complex scalar field $\tau$ parametrizing an $\SL(2)/\SO(2)$ coset.\footnote{The cases $p=3$ and $p=6$ have to be discussed separately.} The bosonic action for the truncated Lorentzian gauged supergravity theories for $0<p<6$ are
\be\label{sugraaction}
S = \f{1}{2\kappa_{p+2}^2}\int\star_{p+2}\left\{ R + \f{3p}{2(p-6)}|\dd \eta|^2 - 2{\cal K}_{\tau\bar{\tau}}|\dd \tau|^2- V\right\}\,,
\ee
where $V$ is the scalar potential and ${\cal K}_{\tau\bar{\tau}}$ the K\"ahler metric obtained from the K\"ahler potential 
\be\label{eq:Kpotdef}
{\cal K} = -\log\left(\frac{\tau-\bar{\tau}}{2}\right) \,,
\ee
The scalar potential can be written in terms of the superpotential
\be\label{superpotential}
{\cal W} = \left\{\begin{array}{ll} 
	-g~\e^{\f{1}{2}\eta}\left(3\tau + (6-p)\rmi\e^{-\f{p}{6-p}\eta}\right) & \text{for $p<3$\,,}\\
	-g~\e^{\f{3(2-p)}{2(6-p)}\eta}\left(3\rmi\e^{\f{p}{6-p}\eta} + (6-p)\tau\right) & \text{for $p>3$\,,}\end{array}\right.
\ee
where $g$ is the gauge coupling constant of the supergravity theory. The scalar potential then reads
\be\label{eq:defVgenp}
V = \f12 \e^{\cal K}\left(\f{6-p}{3p}\left|\partial_\eta {\cal W}\right|^2 + \f14{\cal K}^{\tau\bar{\tau}}D_\tau {\cal W}D_{\bar \tau} \overline{\cal W} - \f{p+1}{2p}\left| {\cal W}\right|^2\right)\,,
\ee
where $D_{\tau} = \partial_{\tau} + \partial_{\tau}{\cal K}$ is the K{\"a}hler covariant derivative. 

For $p=6$ the scalar $\eta$ is not present and the action takes the form 
\be\label{eq:8dsugra}
S = \f{1}{2\kappa_{8}^2}\int\star_8\left\{ R - 2{\cal K}_{\tau\bar{\tau}}|\dd \tau|^2- V\right\}\,,
\ee
where
\be\label{eq:defVp=6}
V = \f12 \e^{\cal K}\left(\f14{\cal K}^{\tau\bar{\tau}}D_\tau {\cal W}D_{\bar \tau} \overline{\cal W} - \f{7}{12}\left| {\cal W}\right|^2\right)\,,
\ee
and
\be\label{eq:Wp=6}
{\cal W} = -3\rmi g\,.
\ee

The actions discussed above are in Lorentzian supergravities and have to be analytically continued. This amounts to taking $\tau$ and its complex conjugate $\tilde{\tau}$ as two independent scalar fields. We should work with two superpotentials, ${\cal W}$ as defined in \eqref{superpotential} and $\widetilde{\cal W}$ given by
\be\label{tildesuperpotential}
\widetilde{\cal W} = \left\{\begin{array}{ll} -g~\e^{\f{1}{2}\eta}\left(3\tilde\tau - (6-p)\rmi\e^{-\f{p}{6-p}\eta}\right) & \text{for $p<3$\,,}\\
	g~\e^{\f{3(2-p)}{2(6-p)}\eta}\left(3\rmi\e^{\f{p}{6-p}\eta} - (6-p)\tilde\tau\right) & \text{for $p>3$\,.}\end{array}\right.
\ee
The scalar potential of the Euclidean theory is obtained by replacing $\overline{\cal W}$ by $\widetilde{\cal W}$ in \eqref{eq:defVgenp}. 

The spherical brane solutions are domain wall backgrounds of the Euclidean supergravity with the following metric
\be\label{eq:genpmetAnsatz}
\dd s_{p+2}^2 = \dd r^2 + {\cal R}^2 \e^{2A} \dd \Omega_{p+1}^2\,.
\ee
The scalar fields and the warp factor $A$ only depend on the radial variable $r$. The constant ${\cal R}$ is the radius of $S^{p+1}$ and is auxiliary since it can be absorbed into a redefinition of the metric function $A$. 

We can now use this Ansatz in the supersymmetry variations of the $(p+2)$-dimensional gauged supergravity theory and look for solutions which preserve 16 supercharges. We find the following system of BPS equations:
\bea
(\eta')^2 &=& \e^{{\cal K}}\left(\f{6-p}{3p}\right)^2(\partial_\eta{\cal W})(\partial_\eta\widetilde{\cal W})\,,\label{lambdaeq}\\
(\eta')(\tau')&=&\e^{{\cal K}}\left(\f{6-p}{12p}\right)\left(\partial_\eta {\cal W}\right){\cal K}^{\tau\tilde{\tau}}D_{\tilde \tau} \widetilde{\cal W} \,,\label{taueq}\\
(\eta')(\tilde\tau')&=&\e^{{\cal K}}\left(\f{6-p}{12p}\right)(\partial_\eta \widetilde{\cal W}){\cal K}^{\tilde{\tau}\tau}D_{\tau} {\cal W} \,,\label{tautildeeq}\\
(\eta')(A' - {\cal R}^{-1}\e^{-A})&=& -\e^{{\cal K}}\left(\f{6-p}{6p^2}\right)(\partial_\eta {\cal W})\widetilde{\cal W}\,,\label{Aeq1}\\
(\eta')(A' + {\cal R}^{-1}\e^{-A})&=& -\e^{{\cal K}}\left(\f{6-p}{6p^2}\right)(\partial_\eta \widetilde{\cal W}){\cal W}\,,\label{Aeq2}
\eea
where ${\cal K}^{\tilde{\tau}\tau}$ is the inverse of the K\"ahler metric. Equations \eqref{lambdaeq}, \eqref{taueq}, and \eqref{tautildeeq} arise from the spin-$\frac{1}{2}$ supersymmetry variations, while \eqref{Aeq1} and \eqref{Aeq2} arise from the spin-$\frac{3}{2}$ variations.

Equations \eqref{Aeq1} and \eqref{Aeq2} lead to a first order differential equation and an algebraic relation for the metric function $A(r)$
\begin{equation}\label{eq:algebraicA}
\e^{A} = \frac{1}{{\cal R}g^2}\frac{2p}{6-p} \frac{\tilde\tau-\tau}{\tilde\tau+\tau}\e^{\frac{2(p-3)}{6-p}\eta}(\eta')\,.
\end{equation}
The BPS equations in \eqref{lambdaeq}-\eqref{Aeq2} are compatible with the second order equations of motion after the analytic continuation.

It is convenient to introduce a new parametrization of the scalar fields as
\be\label{tautoXY}
\begin{array}{ll}
	\tau = \rmi\e^{-\f{p}{6-p}\eta}(X+Y)\,,\quad \tilde\tau = -\rmi\e^{-\f{p}{6-p}\eta}(X-Y)\,,&\text{for $p<3$}\,,\\
	\tau = \rmi\e^{\f{p}{6-p}\eta}(X+Y)\,,\quad \tilde\tau = -\rmi\e^{\f{p}{6-p}\eta}(X-Y)\,,&\text{for $p>3$}\,.
\end{array}
\ee

To find regular solutions of the BPS equations we impose appropriate boundary conditions in the IR. Guided by the physics of the MSYM theory on $S^{p+1}$ we look for solutions in which close to some finite value of the radial coordinate $r \to r_{{\rm IR}}$ the metric is that of $(p+2)$-dimensional flat space in spherical coordinates
\be
\dd s_{p+2}^2 \approx \dd r^2 + (r-r_\text{IR})^2 \dd \Omega_{p+1}^2\,.
\ee

In the UV region, i.e. for large values of $r$, the solution should approach the flat brane domain wall solution for which $X=1$ and $Y=0$. The scalar fields should approach a constant finite value in the IR. These IR values can be found as critical points of the superpotential ${\cal W}$ (or $\widetilde{\cal W}$)
\be
\partial_\eta {\cal W} =  D_\tau {\cal W}=0\,,
\ee
which in terms of the scalars $X,Y$ leads to the solutions in \eqref{eq:XYIR}. The upper and lower sign in \eqref{eq:XYIR} refers to a critical point of ${\cal W}$ and $\widetilde{\cal W}$, respectively. For $p=4$ the critical value of the superpotential is at the UV point $X=1$. 

\subsection{Evaluation of $\kappa_{p+2}$}
\label{app:kappa}

Here we provide a derivation of the Newton constant in $p+2$ dimensions from the one in ten dimensions by employing dimensional reduction.
Consider the metric \eqref{10dmetricgeneral} transformed to Einstein frame and evaluated in the UV, i.e.
\begin{equation}\label{meteinst}
	\dd s_{\rm E}^{2} = g_s^{-1/2}\e^{-\f{(8-p)(p-3)}{4(6-p)}\eta}\left( \dd s_{p+2}^2 + \f{\e^{\f{2(p-3)}{(6-p)}\eta}}{g^2}\dd \Omega^2_{8-p} \right)\,.
\end{equation}
The type II supergravity action is given by
\be
S_{10} = \f{1}{2\kappa_{10}^2}\int\dd^{10}x\sqrt{-g_{(10)}}R_{(10)}+\cdots\,, 
\ee
where the dots represent other terms in the Lagrangian which are not important for the present discussion and $\kappa_{10}^2 = \f{(2\pi\ell_s)^8}{4\pi}$. The $(p+2)$-dimensional supergravity action obtained from this action is
\be 
S_{p+2} = \f{1}{2\kappa_{p+2}^2}\int\dd^{p+2}x\sqrt{-g_{(p+2)}}R_{(p+2)} +\cdots\,, 
\ee 
The goal is now to obtain $\kappa_{p+2}^2$, i.e. the Newton constant on the $(p+2)$-dimensional space. To do this we insert the metric \eqref{meteinst} in the ten-dimensional action and integrate over the internal $(8-p)$-dimensional space. Doing this results in a warp factor which we eliminate by performing the conformal transformation
\be
\tilde{g}_{\mu\nu} = g_s^{-1/2}\e^{-\frac{(8-p)(p-3)}{4(6-p)}\eta}g_{\mu\nu}^{(10)} \,,\qquad \sqrt{-\tilde{g}}\tilde{R} = g_s^{-p/2}\e^{-\frac{p(8-p)(p-3)}{8(6-p)}\eta}\sqrt{-g^{(10)}}R^{{(10)}}\,.  
\ee 
This transformation exactly removes all factors of the function $\eta$ from the internal space and thus we find the unambiguously defined $(p+2)$-dimensional Newton constant 
\be
\f{1}{2\kappa_{p+2}^2} = \f{1}{2\kappa_{10}^2} \frac{V_{8-p}}{g_s^2 g^{8-p}} \,,
\ee
where $V_{n-1} = 2\pi^{n/2}/\Gamma(\tfrac{n}{2})$ is the volume of the unit $n$-sphere. 

\section{Solution for  $d=6$}
\label{appendix:6d}

In this appendix we present an alternative derivation for the free energy and Wilson loop expectation values of $6d$ MSYM 
theory that were obtained in Section \ref{sec:FandWLcases}. For this purpose we start with the saddle-point equation 
(\ref{eom:6d:nonrenorm}) in $d=6-\eps$ dimensions. After renormalization (\ref{lambdarenorm}) of the coupling $\lambda_b$ these equations 
reduce to 
\be
\frac{16\pi^3}{\lambda_\text{QFT}}N\s_i=-3\sum_{j\ne i}(\sigma_i-\sigma_j)\log(\s_i-\s_j)^2\,.
\label{eom:6d:renorm}
\ee
Notice that the r.h.s. of this equation describes repulsive interaction between eigenvalues 
at short distance of the eigenvalues and attractive at large distance. Hence in order to have 
stable distribution of the eigenvalues with large size of the support we should consider $\lambda_\text{QFT}<0$ which is consistent with the conclusions 
of the Section \ref{sec:FandWLcases}. 

\begin{figure}[!tbp]
	\centering
 \includegraphics[width=0.65\linewidth]{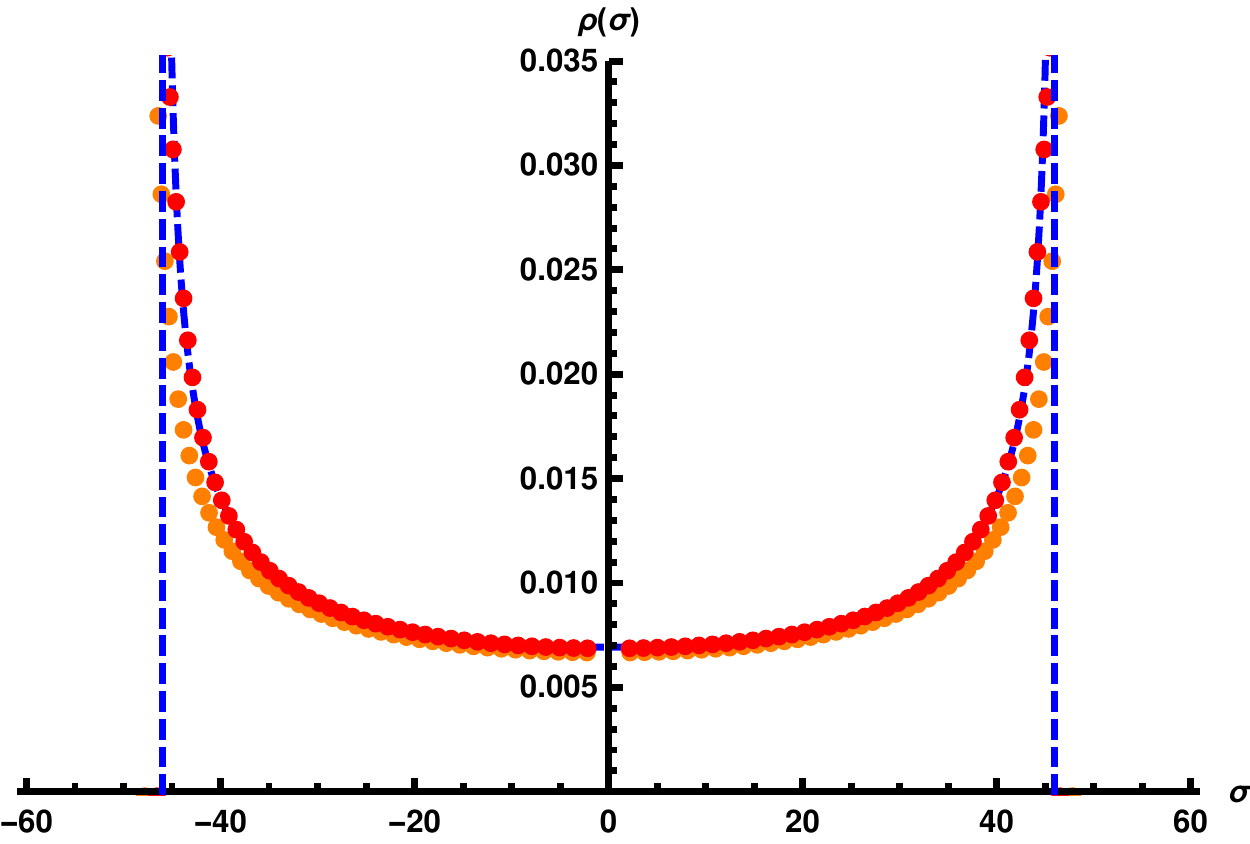}
 \caption{The eigenvalue distribution of $6d$ MSYM for $N=100$ and $\lambda_\text{QFT}=-20$. In particular the \textit{orange  dots} correspond
         to the numerical solution of (\ref{saddlept})  at $\lambda_b=0.42$ and $d=5.995$ ($\eps=0.005$). The latter parameters correspond
	 to $\lambda_\text{QFT}=-20$ according to (\ref{lambdarenorm}). The  \textit{red dots} 
	 in turn correspond to the numerical solution of (\ref{eom:6d:renorm}) at $N=100$ and $\lambda_\text{QFT}=-20$. Finally the dashed blue
	 line shows the eigenvalue density (\ref{dens:6d}) with the endpoint position $b$ given by (\ref{endpt:6d:appr}).
	}
	\label{pic:6d}
\end{figure}

In the continuous limit we as usually introduce eigenvalue density according to (\ref{densdelta}) and rewrite saddle-point equation (\ref{eom:6d:renorm}) as 
the following integral equation
\be
-\frac{16\pi^3}{3\lambda_\text{QFT}}\s=\int\limits_{-b}^b d\s'\rho(\s') (\s-\s')\log(\s-\s')^2\,.
\label{eom:6d:cont}
\ee
This equation has already appeared before in the context of $4d$ ${\cal N}=2$ theories in \cite{Russo:2012ay,Russo:2013qaa}. 
To solve it we should differentiate it twice w.r.t. 
$\s$ in order to obtain 
\be
\int\limits_{-b}^b d\s' \frac{\rho(\s')}{\s-\s'}=0\,,
\ee
which is the standard singular integral equation with Cauchy kernel. This equation has the following unbounded normalizable solution
\be
\rho(\s)=\frac{1}{\pi \sqrt{b^2-\s^2}}\,,
\label{dens:6d}
\ee
In order to define position of the support endpoint $b$ we can use the following integral: 
\be
\int\limits_{-b}^b d\s'\frac{(\s-\s')\log(\s-\s')^2}{\pi\sqrt{b^2-\s'^2}}=2\s\log\left(\frac{be}{2}  \right)\,.
\ee
Comparison with (\ref{eom:6d:cont}) immediately gives 
\be
b=2\exp\left(-\frac{8\pi^3}{3\lambda_\text{QFT}}-1\right)\,,
\label{endpt:6d:appr}
\ee
which precisely reproduces expression (\ref{b6}) we have obtained previously considering $\eps\to 0$ of general expression (\ref{bres}). 
It is also worth noticing that the eigenvalue density (\ref{dens:6d}) which  solves (\ref{eom:6d:cont})  is consistent with the $\eps\to 0$ 
limit of the general expression (\ref{eigdens}) provided we also use coupling renormalization (\ref{lambdarenorm}). On Fig.\ref{pic:6d} 
we also compare numerical solutions of equations (\ref{eom:6d:renorm}), (\ref{saddlept}) and analytical solution (\ref{dens:6d}). As we see solutions 
to the equation (\ref{saddlept}) with full kernel agrees with the solution of (\ref{eom:6d:renorm}) when $d$ is close to $6$. Also the solution 
(\ref{dens:6d}) describes both numerical solutions very well. 

Finally to find the free energy instead of substituting eigenvalue density (\ref{dens:6d}) into free energy functional we notice the following identity
\be
\frac{1}{N^2}\frac{\partial F}{\partial\lambda_r}=-\frac{8\pi^3}{\lambda_\text{QFT}^2}\int\limits_{-b}^b \rho(\s)\s^2=
-\frac{16\pi^3}{\lambda_\text{QFT}^2}\e^{-\frac{16\pi^3}{3\lambda_\text{QFT}}-2}
\ee
Integrating this identity we easily obtain 
\be
\frac{F}{N^2}=-3\e^{-\frac{16\pi^3}{3\lambda_\text{QFT}}-2}\,,
\ee
which exactly reproduces expression in (\ref{F6r}). Wilson loop we will obtain from (\ref{dens:6d}) and (\ref{endpt:6d:appr}) will obviously also 
reproduce previously obtained result (\ref{WL6}).

\section{Solutions for $d=7$}
\label{appendix:7d}

\subsection{An alternative derivation for the eigenvalue density}
\label{appendix:7dalt}

In this part of the appendix we present an alternative way to analyze the matrix model 
\eqref{partfun} for the $d=7$ case. For this purpose we have to solve the saddle point 
equation \eqref{eom}
\be
\frac{C_1}{\lambda_\text{QFT}}\,\s=C_2\int_{-b}^{b}d\s'\rho(\s')|\s-\s'|^2\sign(\s-\s')\,
\label{eom:7d}
\ee
with the regularized 't Hooft coupling $\lambda_\text{QFT}$ which we assume is small and negative such that the eigenvalues are in general widely separated.
The integral equation in \eqref{eom:7d} is closely related to the saddle point equation of 
the  matrix model for five-dimensional SYM in the decompactification limit, i.e.
when the radius $\mathcal{R}$ of $S^5$ is taken to infinity.  A detailed analysis of this matrix model can be found 
in \cite{Nedelin:2015mta}. 

To solve \eqref{eom:7d} we  differentiate both sides of the equation twice with respect to $\s$. This leads to 
the simple equation
\be
\int_{-b}^{b}d\s'\rho(\s')\sign(\s-\s')=\int_{-b}^{\s}d\s'\rho(\s')-\int_{\s}^{b}d\s'\rho(\s')=0\,.
\ee
This equation should be satisfied for any $\s$ on the support, but this is possible only if the
eigenvalue density $\rho(\s)$ is zero everywhere except at the support endpoints, $\pm b$. Hence, we can assume the following form for the 
solution,
\be
\rho(\s)=\frac{1}{2}\left( \delta\left( \sigma+b \right)+\delta\left( \sigma-b \right) \right)\,,
\label{density:7d}
\ee
where the  factor of $1/2$ is introduced to normalize the eigenvalue density. The endpoints of the distribution can then be found by substituting the density \eqref{density:7d} back into the integral equation \eqref{eom:7d}.  This then results in the simple algebraic equation 
\be
C_1\lambda_\text{QFT}^{-1}\sigma=\frac{1}{2}C_2 \left[ \left( \sigma+b \right)^2-\left( \sigma-b \right)^2 \right]\,,
\ee
which, using \eqref{saddlept} and \eqref{C2def}, leads to
\be
b=\frac{1}{2\lambda}\frac{C_1}{C_2}=-\frac{2\pi^3}{\lambda_\text{QFT}}\,.
\label{b:7d}
\ee
The final expression agrees  with \eqref{b7} determined from the general expression. 
One can also easily obtain the free energy in \eqref{F7} using the distribution in 
\eqref{density:7d} and the value of $b$ in \eqref{b:7d}.

Note that we can also derive the $\delta$-function behavior in \eqref{density:7d} directly
from the expression for the density in \eqref{eigdens}. 
If we let $d=7-\eps$
then it is straightforward to see that the density approaches zero in the limit $\eps\to0$, everywhere except at $\s=\pm b$. 

\subsection{Numerical solutions at weak negative coupling}
\label{appendix:negcoup}

\begin{figure}[!tbp]
	\centering
	\subfigure[]{\includegraphics[width=0.45\linewidth]{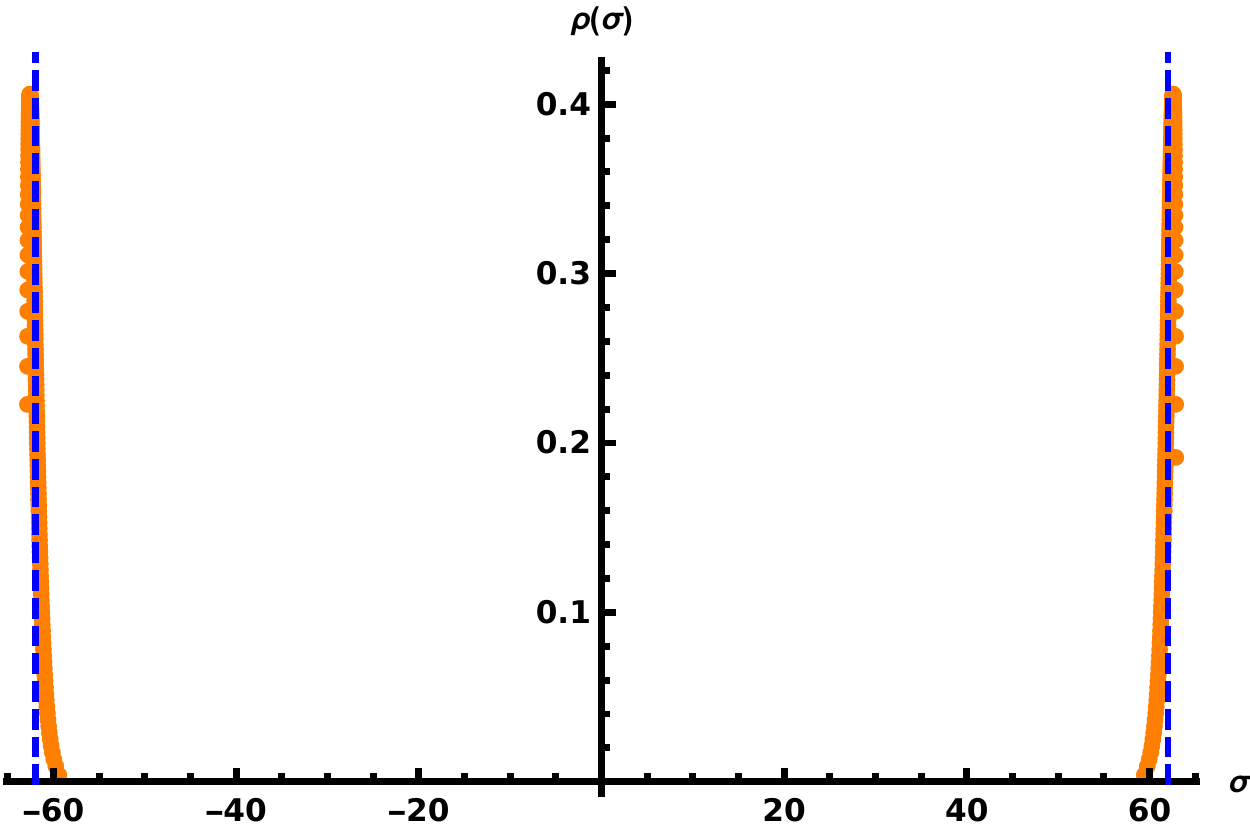}}
	\subfigure[]{\includegraphics[width=0.45\linewidth]{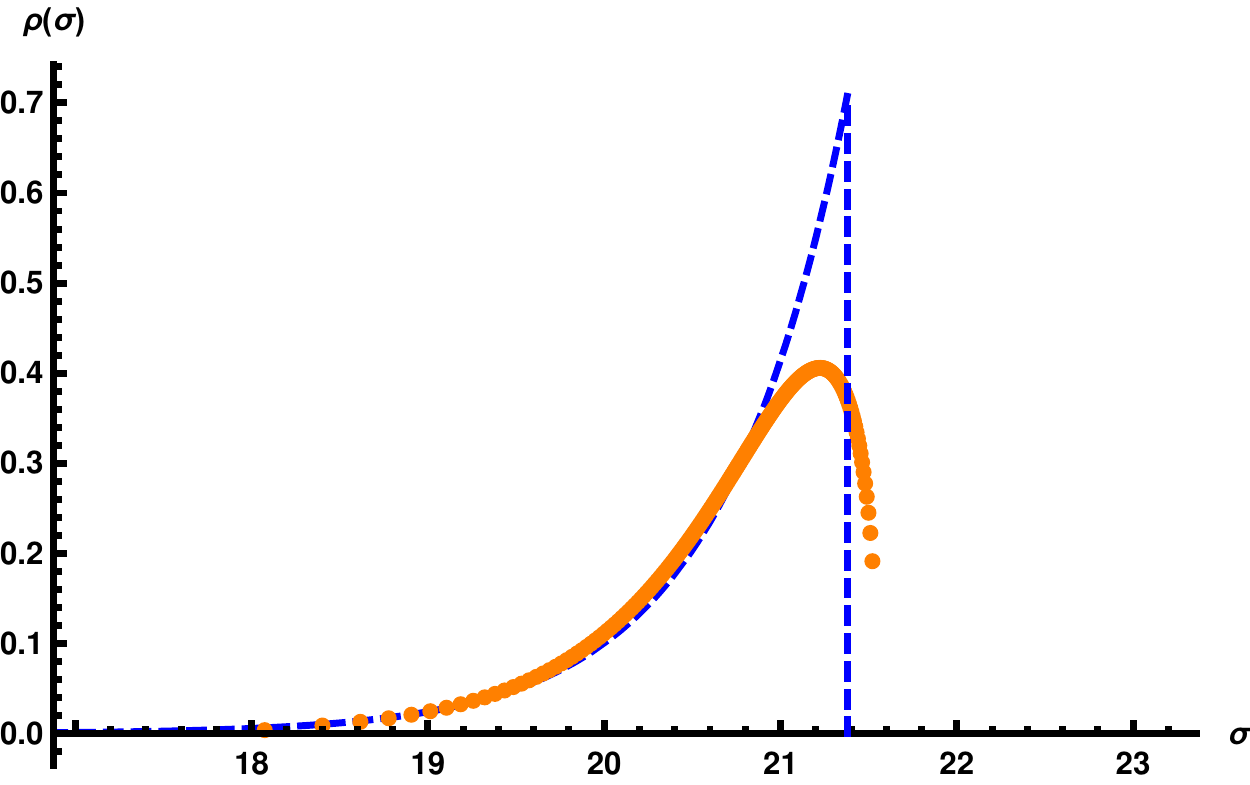}}
	\caption{Eigenvalue distribution of $7d$ MSYM for $N=400$ and $\lambda_\text{QFT}=-1$ and $\lambda_\text{QFT}=-3$. In the \textit{left panel} we compare
		numerical results for $\lambda_\text{QFT}=-1$ against the eigenvalue density \eqref{density:7d}, while on the \textit{right panel} we
		compare numerics for $\lambda_\text{QFT}=-3$ against the analytical solution \eqref{dens7}.
		The latter one takes into account the finite width of the eigenvalue support. 
	}
	\label{pic:7d:2}
\end{figure}

In this part of the appendix we analyze numerically the solution to (\ref{saddlept}) for $d=7$ and a weak negative renormalized 't Hooft constant.  
Here we solve a ``heat equation'' numerically\footnote{See \cite{Nedelin:2015mta} for a more detailed 
explanation of the numerical techniques.}, which at large ``times'' approaches asymptotically  the
solution to the saddle point equation below in \eqref{eom:7d:large:separ}.
However, we also assume that  the solution is symmetric around the origin, i.e.
for each eigenvalue $\sigma_i$  there is another eigenvalue $-\sigma_i$. As can be seen from  
Figure~\ref{pic:7d:2} (a), which compares the numerical and analytical results, the solution
in \eqref{density:7d} indeed reproduces the behavior of the eigenvalue distribution
at weak negative coupling. Notice that the graph  is for {$\lambda_\text{QFT}=-1$} which is not
very small. Our solution  works whenever $\frac{|\lambda_\text{QFT}|}{ {4}\pi^3}\ll 1$, which 
obviously holds for {$\lambda_\text{QFT}=-1$}.

For  $-\lambda_\text{QFT}^{-1}\gg \frac{1}{4\pi^3}$ the eigenvalue distribution separates into two widely separated peaks 
according to \eqref{density:7d} with distance $\frac{4\pi^3}{|\lambda_\text{QFT}|}$ between them.  
However if we include subleading terms in the kernel we can argue that the peaks are actually humps with a width of order 1. 
Including the next term, $G_{16}^{(7)}(\s)$ in \eqref{kerneld7} has the expansion
\be\label{G167approx}
G_{16}^{(7)}(\s)= 2\pi(1-\s^2)\sign(\s)+{\rm O}(\e^{-\pi|\s|})\,.
\ee
At this order in the approximation  there is a repulsive force at short distance which smears the peaks into finite 
size humps. To estimate the size of these humps we note that the eigenvalue distribution is even about $\sigma=0$.  Hence, assuming that $N$ is even, we can express the eigenvalues as
\bea
\sigma_i&=&-\sigma_0-\delta\sigma_i\qquad\ \  1\le i\le N/2\,,\nonumber\\
\sigma_{i}&=&\sigma_0+\delta\sigma_{N-i}\qquad N/2+1\le i\le N\,.
\eea
Here we assume that $\sigma_0\gg|\delta\sigma_i|$ and
\be\label{sumeq}
\sum_{i=1}^{N/2}\delta\sigma_i=0\,,
\ee
to keep the center of mass of the eigenvalues on each hump fixed at $\pm\s_0$. The equation of motion in \eqref{saddlept} for the positive eigenvalues can then be well approximated as
\be\label{7deomapp}
\frac{4 \pi^3 N}{\lambda_\text{QFT}}(\sigma_0+\delta\sigma_i)=\sum\limits_{j\neq i}^{N/2}\left[ 1-\sigma_{ij}^2 \right]\coth\left( \pi\,\sigma_{ij} \right)
+\sum_{j=1}^{N/2}(1-(2\sigma_0+\delta\sigma_i+\delta\sigma_j)^2)\,.
\ee
Setting $\sigma_0=-\frac{2\pi^3}{\lambda}$, (\ref{7deomapp}) reduces to
\be\label{7dred}
-1=\frac{2}{N}\sum\limits_{j\neq i}^{N/2}\left(\left[ 1-\sigma_{ij}^2 \right]\coth\left( \pi\,\sigma_{ij} \right)-\sigma_{ij}^2\right)\,,
\ee
where the condition on the sum in \eqref{sumeq} is also imposed.  This last equation has no $\lambda$ dependence and we can expect that the $\sigma_i$ range over a size of order 1.  This can be confirmed numerically as can be seen in Figure~\ref{pic:7d:2} (b). 

In Figure~\ref{pic:7d:2} (b) we can see an exponential fall-off of the humps. We can capture this behavior using
the expression \eqref{G167approx} for the kernel. In this case the continuous limit of \eqref{saddlept} can be written as 
\be
\frac{ {4}\pi^3}{\lambda_\text{QFT}}\,\s=\int_{-b}^\s (1-(\s-\s')^2)\rho(\s')d\s'-\int^{b}_\s (1-(\s-\s')^2)\rho(\s')d\s'\,.
\ee
Taking three derivatives with respect to $\s$ on both sides of the equation gives
\be
2\,\rho''(\s)-4\rho(\s)=0\,,
\ee
hence
\be\label{dens7}
\rho(\s)=\frac{k}{\sqrt{2}}\cosh\left(\sqrt{2}\,\s\right)\,,
\ee
with the constraints
\bea\label{C1}
&&1=\int_{-b}^b \frac{k}{\sqrt{2}}\cosh\left({\sqrt{2}\,\s}\right)d\s=k\sinh\left({\sqrt{2}\,b}\right)\,,\\
\label{C2}&&\frac{ {4}\pi^3}{\lambda_\text{QFT}}=\sqrt{2}k\left(\cosh\left({\sqrt{2}\,b}\right)-{\sqrt{2}\, b}\sinh\left({\sqrt{2}\,b}\right)\right)\,.
\eea
Using \eqref{C1} we can rewrite \eqref{C2} as
\be\label{C3}
\sinh\left(\sqrt{1+k^2}-t\right)=\frac{1}{k}\,,
\ee
where $t\equiv{ {2}\sqrt{2}\pi^3}/{\lambda_\text{QFT}}$. For a given $\lambda_\text{QFT}$ one can solve \eqref{C3}
for $k$ numerically, and thus determine $\rho(\s)$ in \eqref{dens7}.
The dashed line in Figure~\ref{pic:7d:2} (b) shows this density at $\lambda_\text{QFT}=-3$.

\subsection{Solutions at weak negative coupling and finite $N$}
\label{appendix:finiteN}

In this part of the appendix we consider $d=7$ solutions at finite $N$.
For small negative regularized coupling we can use the approximate  equations of motion in \eqref{eom:strong}, which for $d=7$ are
\be
-\frac{ {4} \pi^3 N}{\lambda_\text{QFT}}\sigma_i=\sum\limits_{j\neq i}\sigma_{ij}^2 \mathrm{sign}\left( \sigma_{ij} \right)\,.
\label{eom:7d:large:separ}
\ee
Assuming that $N$ is even, the solution  that corresponds to the large $N$ solution  in the previous subsection has $N/2$
eigenvalues at $\sigma_+=b_7=-\frac{2\pi^3}{\lambda_\text{QFT}}$ and $N/2$ at $\s_-=-\sigma_+$.    However, if we put $2M$
eigenvalues at $\s=0$, then we can also satisfy  \eqref{eom:7d:large:separ} if we place $N/2-M$ eigenvalues at
$\s_+=-\frac{2\pi^3}{\lambda_\text{QFT}}\frac{N}{N-M}$ and the same number at $\s_-=-\sigma_+$ 
\footnote{There are still other solutions, {\it e.\,g.} one can have an unequal number 
of eigenvalues at $\s_+$ and $\s_-$, in which case $\s_-\ne-\s_+$.}.  We have ignored 
short range interactions here, but as is shown in the previous subsection they only spread the eigenvalues an order 1 distance  from the peaks.

The free energy for these more general solutions is given by
\be\label{Ffinite}
F=\frac{16\pi^{10}N^2}{3\lambda_\text{QFT}^3}\left(1-\frac{M^2}{(N-M)^2}\right)\,,
\ee
demonstrating that the free energy increases with increasing $M$ since $\lambda_\text{QFT}<0$.  Assuming that $\lambda_\text{QFT}$
is also small \eqref{Ffinite} shows that the solutions with nonzero $M$ are heavily suppressed.  If $N$ is odd then $M$ is replaced with $M+1/2$ in \eqref{Ffinite}.

The quadratic fluctuations about the lowest energy solution are
\bea
\delta F&=&\frac{4\pi^4N}{\lambda_\text{QFT}}\sum_i(\delta\s_i)^2+2\pi\sum_{j\ne i}|\s_i-\s_j|(\delta\s_i)^2-2\pi\sum_{j\ne i}|\s_i-\s_j|\delta\s_i\delta\s_j\nn\\
&=&\frac{4\pi^4}{\lambda_\text{QFT}}\left(\sum_i^{N/2}\delta\s^{(+)}_i\right)\left(\sum_i^{N/2}\delta\s^{(-)}_i\right)\,,
\eea
where $\delta\s_i^{(+)}$ are the fluctuations of the eigenvalues at $\s_+$ and $\delta\s_i^{(-)}$ are the fluctuations
of the eigenvalues at $\s_-$.  Hence, to quadratic order there is a tachyonic mode corresponding to the overall center
of mass motion, which is not present for $SU(N)$, and a positive mode corresponding to the average of the left and 
right eigenvalues moving in the opposite direction.  This latter mode has a large positive coefficient and thus is
sharply suppressed.  All other modes are zero modes.  

The zero modes are not exact as there are nonzero cubic terms.  Since the center of mass modes are either
removed or suppressed, we can assume that $\sum_i\delta\s_i^{(+)}=\sum_i\delta\s_i^{(-)}=0$.  Then the fluctuations of the free energy are
\bea
\delta F&=&\frac{\pi}{3}\sum_{i,j}^{N/2}\left(|\delta\s_i^{(+)}-\delta\s_j^{(+)}|^3+|\delta\s_i^{(-)}-\delta\s_j^{(-)}|^3+2(\delta\s_i^{(+)}-\delta\s_j^{(-)})^3\right)\nn\\
&=&\frac{\pi}{3}\left(N\sum_i^{N/2}(\delta\s_i^{(+)})^3+\sum_{i,j}^{N/2}|\delta\s_i^{(+)}-\delta\s_j^{(+)}|^3\right) \nn\\
&&\qquad\qquad\qquad+ \frac{\pi}{3}\left(-N\sum_i^{N/2}(\delta\s_i^{(-)})^3+\sum_{i,j}^{N/2}|\delta\s_i^{(-)}-\delta\s_j^{(-)}|^3\right)\,,
\eea
where we see that right and left fluctuations decouple from each other.  Note that these fluctuations are of order 1 and independent
of $\lambda_\text{QFT}$, hence their only effect is to shift the free energy by an unimportant constant and can be ignored even for small $N$.

\bibliographystyle{utphys}
\bibliography{refs}

\end{document}